\documentclass{aa}
\usepackage{graphicx}
\usepackage{txfonts}
\usepackage{natbib}
\bibpunct{(}{)}{;}{a}{}{,}  

\begin{document}

\newcommand{\zabs}{z_{\rm abs}}
\newcommand{\dla}{damped Lyman-$\alpha$}
\newcommand{\DLA}{Damped Lyman-$\alpha$}
\newcommand{\lya}{Ly-$\alpha$}
\newcommand{\lyb}{Ly-$\beta$}
\newcommand{\HI}{H\,{\sc i}}
\newcommand{\CI}{C\,{\sc i}}
\newcommand{\CII}{C\,{\sc ii}}
\newcommand{\CIV}{C\,{\sc iv}} 
\newcommand{\PII}{P\,{\sc ii}}
\newcommand{\SII}{S\,{\sc ii}}
\newcommand{\SiII}{Si\,{\sc ii}}
\newcommand{\SiIV}{Si\,{\sc iv}}
\newcommand{\FeII}{Fe\,{\sc ii}}
\newcommand{\ZnII}{Zn\,{\sc ii}}
\newcommand{\NiII}{Ni\,{\sc ii}}
\newcommand{\MgI}{Mg\,{\sc i}}
\newcommand{\MgII}{Mg\,{\sc ii}}
\newcommand{\ClI}{Cl\,{\sc i}}
\newcommand{\ClII}{Cl\,{\sc ii}}



\title{Physical conditions in the neutral interstellar medium at $z =
  2.43$ toward Q\,2348$-$011\thanks{Based on observations carried
    out at the European Southern Observatory (ESO) under prog. ID
    No. 072.A-0346 with the UVES spectrograph installed at the Very
    Large Telescope (VLT) Unit 2, Kueyen, on Cerro Paranal, Chile.}}

\titlerunning{Physical conditions in the neutral ISM at $\zabs=2.43$ toward Q\,2348$-$011}

\author{P. Noterdaeme\inst{1} \and P. Petitjean\inst{2,3} \and
  R. Srianand\inst{4} \and C. Ledoux\inst{1} \and F. Le Petit\inst{5}}

\institute{European Southern Observatory, Alonso de C\'ordova 3107, Casilla 19001, 
Vitacura, Santiago, Chile\\
\email{pnoterda@eso.org, cledoux@eso.org}
\and
Institut d'Astrophysique de Paris, CNRS - Universit\'e Pierre et Marie
Curie, 98bis Boulevard Arago, 75014, Paris, France\\
\email{petitjean@iap.fr}
\and
LERMA, Observatoire de Paris, 61 Avenue de l'Observatoire, 75014, Paris, France
\and
IUCAA, Post Bag 4, Ganesh Khind, Pune 411\,007, India\\
\email{anand@iucaa.ernet.in}
\and
LUTH, Observatoire de Paris, 61 Avenue de l'Observatoire, 75014, Paris, France\\
\email{franck.lepetit@obspm.fr}
}

\date{Received ; accepted}

\abstract
{}
{We aim at deriving the physical conditions in the neutral gas associated with \dla\ systems
using observation and analysis of H$_2$ and C~{\sc i} absorptions.}
{We obtained a high-resolution VLT-UVES spectrum of the quasar Q\,2348$-$011
over a wavelength range that covers most of the prominent metal and
molecular absorption lines from
the $\log N$(H~{\sc i})~=~20.50$\pm$0.10 \dla\ system at $\zabs=2.4263$.  
We detected H$_2$ in this system and measured column densities of H$_2$, \CI, \CI$^*$, 
\CI$^{**},$ \SiII, \PII, \SII, \FeII, and \NiII. 
From the column density ratios and, in 
particular, the relative populations of H$_2$ rotational and \CI\ fine-structure levels, we derived the 
physical conditions in the gas (relative abundances, dust-depletion, particle density, kinetic
temperature, and ionising flux) and discuss physical conditions in the neutral phase. 
}
{Molecular hydrogen was detected in seven components in the first four
  rotational levels (J~=~0-3) of the vibrational ground state. Absorption lines of H$_2$ J~=~4 (resp. J~=~5) rotational levels are
detected in six (resp. two) of 
these components. This leads to a total molecular fraction of  $\log f$~$\simeq
-1.69^{+0.37}_{-0.58}$.
Fourteen components are needed to reproduce the metal-line profiles. 
The overall metallicity is found to be $-$0.80, $-$0.62, $-$1.17$\pm$0.10 for,
respectively, [Si/H], [S/H] and [Fe/H].
We confirm the earlier findings that there is a correlation between $\log N$(\FeII)/$N$(\SII)
and $\log N$(\SiII)/$N$(\SII) from different components indicative of a dust-depletion
pattern. Surprisingly, however, the depletion of metals onto dust in the H$_2$ components
is not large 
in this system: 
[Fe/S]~=~$-$0.8 to $-0.1$.

The gas in H$_2$-bearing components is found to be 
cold but still hotter than similar gas in our Galaxy ($T > 130$~K,
instead of typically 80~K) and 
dense ($n \sim 100-200$~cm$^{-3}$). 
There is an anti-correlation (R~$=-0.97$) between 
the logarithm of the photo-absorption rate, $\log \beta_0$, and $\log N$(H$_2$)/$N$(\CI) derived for each
H$_2$ component. We show that this is mostly due to shielding effects and
imply that the photo-absorption rate $\beta_0$ is a good indicator of the physical
conditions in the gas. We find that the gas is immersed in an 
intense UV field, about one 
order of magnitude higher than in the solar vicinity. 
These results suggest that the gas in H$_2$-bearing DLAs is clumpy, and
star-formation occurs in the associated object.
}
{}


\maketitle

\section{Introduction}
High-redshift \dla\ systems (DLAs) detected in absorption
in QSO spectra are characterised by their large neutral hydrogen
column densities, $N($\HI$)\ge 2\times 10^{20}$~cm$^{-2}$, similar
to what is measured through local spiral disks \citep{Wolfe86}. The corresponding absorbing 
clouds are believed to be the reservoir of neutral hydrogen in the 
Universe \citep[e.g.,][]{Peroux03, Prochaska05}.
Though observational studies of DLAs have been pursued over more than
two decades \citep[see][for a recent review]{Wolfe05}, 
important questions are still unanswered, such as (i) the amount of in-situ star-formation 
activity in DLAs, (ii) the connection between observed abundance ratios and the dust 
content, and (iii) how severe is the bias due to dust obscuration in current DLA samples. 

One way to tackle these questions is to derive the physical conditions
that prevail in the absorbing gas by studying the excitation
of molecular and atomic species together. 
Since UV-lines are redshifted into the optical,
a single 
VLT-UVES spectrum allows one to simultaneously observe a large number 
of absorption lines of important species 
at high spectral resolution and with a good signal-to-noise ratio.
%
In the case of the Galactic interstellar medium (ISM),
only a few lines of sight have been observed in such detail.
\par \noindent
%

%
%
\par
Although molecular hydrogen is the most abundant molecule in the
Universe, it is very difficult to detect directly.  
The molecule has a rich absorption spectrum in the UV, and spacecrafts have been
used to detect the absorption lines from clouds located in our Galaxy 
in front of bright background stars.
At \HI\ column densities as high as those measured in DLAs or sub-DLAs
($N$(\HI)~$>$~19.5~cm$^{-2}$), H$_2$ molecules are conspicuous in our
Galaxy: sightlines 
with $\log N($\HI$)>21$ usually have $\log N($H$_2)>19$ \citep[see][]{Savage77, Jenkins79}.
%
More recently, observations with the Far Ultraviolet Spectroscopic
Explorer (FUSE) have derived the physical conditions in the diffuse molecular gas in the Galaxy 
but also the Magellanic clouds \citep{Tumlinson02} or high Galactic latitude lines of
  sight \citep{Richter03,Wakker06}. 
  In the Magellanic clouds, molecular fractions and kinetic temperatures 
are reported to be, respectively, lower and higher than in the local ISM. 
This is probably a consequence of higher UV ambient flux and lower
dust content.
  In intermediate velocity clouds located in the Galactic halo, H$_2$ is found to arise in
  cold ($T \la 140$~K), dense ($n_{\rm H} \sim 30$~cm$^{-3}$), and clumpy
  ($D\sim0.1$~pc) clouds \citep{Richter03}, embedded in a relatively low UV flux.
\par
At high redshift, UV-lines are redshifted in the optical so ground-based
telescopes can be used. Quasars
\citep[e.g.,][]{Levshakov92,Ge99,Ledoux03} but also recently 
$\gamma$-ray burst afterglows \citep{Fynbo06} are used as background sources.
The amount of H$_2$ molecules seen in DLAs is much less than what is observed
in our Galaxy.
\citet{Ledoux03} find that about 15\% of DLAs show detectable H$_2$ with 
molecular fraction in the range $-4 < \log f < -1$. This fraction increases
with metallicity and reaches 50\% for [X/H]~$>$~$-0.7$ \citep{Petitjean06}.
The fact that H$_2$ is less conspicuous at high redshift can be explained 
as a consequence of lower metal content and therefore lower dust content, together with 
higher ambient UV flux. It could be also that the covering factor of the
diffuse molecular gas is low.
\par Detailed analysis of H$_2$-bearing DLAs has been pursued
in some cases. \citet{Cui05} recently derived a surprisingly low hydrogen density, 
$n_{\rm H}\approx 0.2$~cm$^{-3}$, using R~$\sim$~30,000 STIS data from the HST, and 
an electron temperature of $T_{\rm e} \approx 140$~K in the H$_2$ system at 
$\zabs=1.776$ toward Q\,1331$+$170. Other authors, however, usually derive higher densities.
By studying the spectrum of the $\zabs\simeq2.811$ DLA 
toward PKS\,0528$-$250, \citet{Srianand98}
derived a kinetic temperature of $T \sim 200$~K and a density
of $n \sim 1000$~cm$^{-3}$. They suggest that the ratio $N($H$_2)/N($\CI$)$ 
could be a good indicator of the physical conditions in the gas. \citet{Petitjean02} 
derived gas pressure, particle density, excitation temperatures, and
ambient UV fluxes for four H$_2$ components at $\zabs\simeq 1.973$ toward
Q\,0013$-$004. They show that, whenever H$_2$ is detected, the particle
density is high ($n_{\rm H} \sim 30 - 200$~cm$^{-3}$) and the kinetic temperature is
low ($T \sim 100$~K). In addition, the ambient UV radiation field is
found to be highly inhomogeneous.
\citet{Ledoux02} find high particle densities ($n \sim 30 -
  400$~cm$^{-3}$) and moderate UV radiation field in a DLA at
  $\zabs=1.962$. They conclude that physical conditions play an important role in governing the molecular
fraction in DLAs.
  \citet{Reimers03} also found a high particle density ($n \sim
  100$~cm$^{-3}$) in a DLA at $\zabs=1.150$. From the high derived UV
  radiation field, they suggest there is ongoing
star-formation activity close to the absorbing cloud.
\citet{Srianand05} use the 33 systems from
the high-resolution (R~$\sim$~45,000) UVES survey by \citet{Ledoux03} -- among which 8 show 
detectable H$_2$ absorption lines -- to compare observations and molecular cloud models.
They conclude that the mean kinetic temperature in H$_2$-bearing DLAs is 
higher ($\sim 150$~K) than in the local ISM ($\sim 80$~K), 
the UV flux is close to or higher than the mean UV flux 
in the ISM of the Galaxy, and the particle densities ($\sim$~10$-$200~cm$^{-3}$) and photo-ionisation 
rate are similar to the ones in the Galactic cold neutral medium.  More recently, gas temperature, 
$T \sim 90 - 180$~K, and density, $n_{\rm H}\leq 50$~cm$^{-3}$, were derived in a DLA
at $\zabs\simeq 4.224$, which corresponds to the highest redshift at which
this kind of study has been done up to now \citep{Ledoux06b}. 


\par
We recently reported the detection of H$_2$ in
the $\zabs =2.4263$ DLA toward Q\,2348$-$011 \citep{Petitjean06}. 
We present here a detailed analysis of the physical conditions in the gas.
This system has one of the highest molecular fractions ever measured
at high redshift, and seven H$_2$ components are detected in which physical 
conditions can be derived.
This is the first time that a comparison of physical conditions can
be made in such a large number of components in a single absorption
system. 
The paper is organised as follows. In Sect.~2, we describe observations 
and data reduction. In Sect.~3, we give details
  on the data and present the fits to the H$_2$, \CI\ and
  low-ionisation metal absorption lines. In Sect.~4, we discuss the
  relative populations of H$_2$ rotational levels, and 
C~{\sc i} fine-structure levels, deriving information on the physical
  state of the gas. Finally, we discuss our findings and draw our conclusions in Sect.~5.

\section{Observations}
The quasar 
Q\,2348$-$011 was observed with the Ultraviolet 
and Visible Echelle Spectrograph \citep[UVES; ][]{Dekker00} 
mounted on the ESO Kueyen VLT-UT2 8.2 m telescope 
on Cerro Paranal in Chile in visitor mode on October 29 and 30,
2003, under good seeing conditions ($FWHM < 0.8$~arcsec)
and at low airmass ($AM < 1.2$).
Three 5400\,s exposures were taken with Dichroic \#1 simultaneously
with the blue and red arms.
We used a 1.0 arcsec slit width and a pixel binning of $2 \times 2$, leading
to a resolving power of $R$~$\simeq$~47,000.
The data were reduced using the UVES pipeline \citep{Ballester00},
which is available in the context of the ESO MIDAS data reduction
system.
The main characteristics of the pipeline are to
perform a precise inter-order background subtraction, especially for master
flat fields, and to allow for an optimal extraction of the object
signal rejecting cosmic rays and performing sky subtraction at the same time.
The pipeline products were checked step-by-step.

The wavelength scale of the reduced spectra was then converted to vacuum-heliocentric 
values, and
the portions of the spectrum corresponding to different
arms were rebinned to a constant wavelength step. 
Individual 1-D exposures were scaled, weighted, and combined. 
No other rebin was performed during the analysis of the whole spectrum.
Standard Voigt-profile fitting methods were used for the analysis to determine column
densities using the wavelengths and oscillator strengths 
compiled in Table~1 of \citet{Ledoux03} 
for metal ions and the oscillator
strengths given by \citet{Morton76} for H$_2$.
We adopted the Solar abundances given by \citet{Morton03} based on the
meteoritic data from \citet{Grevesse02}.

\section{The DLA at $\zabs=2.4263$ toward Q\,2348$-$011}
There are two DLAs at $\zabs=2.42630$ and $\zabs=2.61473$ 
toward Q\,2348$-$011. These DLAs were discovered as strong
Lyman-$\alpha$ absorptions in a low spectral-resolution ($\sim 5 \AA$) spectrum
by \citet{Sargent89}, who also observed
strong metal lines associated to the $\zabs=2.4263$ DLA.
From Voigt-profile fitting to the \dla\ line, we find
that the total neutral hydrogen column density of the system at $\zabs=2.4263$ 
is $\log N$(\HI) = $20.50\pm0.10$ \citep{Ledoux06a}.
A simultaneous Voigt-profile fitting of the Lyman-$\alpha$ and
Lyman-$\beta$ lines was also performed for the system at
$\zabs=2.6147$. Note that the $\zabs=2.6147$ Lyman-$\beta$ line
falls in the wavelength range where  H$_2$ lines at $\zabs=2.4263$
are detected (see Fig.~\ref{q2348_H2}). The total neutral hydrogen column density for this
system is  $\log N$(\HI$)$ = $21.30\pm0.08$ \citep{Ledoux06a}. No molecular hydrogen is
detected at this redshift down to $\log N$(H$_2$)~$<$~13.7
(considering rotational levels J~=~0 and J~=~1), or $\log f<$~$-$7.2.

\subsection{Molecular content}
The H$_2$ absorption lines at $\zabs=2.4263$ are strong and numerous (see Fig.~\ref{q2348_H2}). Although
they are all redshifted in the Lyman-$\alpha$ forest, the resolving
power of our data ($R \simeq 47,000$) is high enough to allow unambiguous
detection and accurate determination of the line parameters. 

\subsubsection{Overall Fit}
\begin{figure*}[!ht]
 \begin{center}
 \includegraphics[width=\hsize]{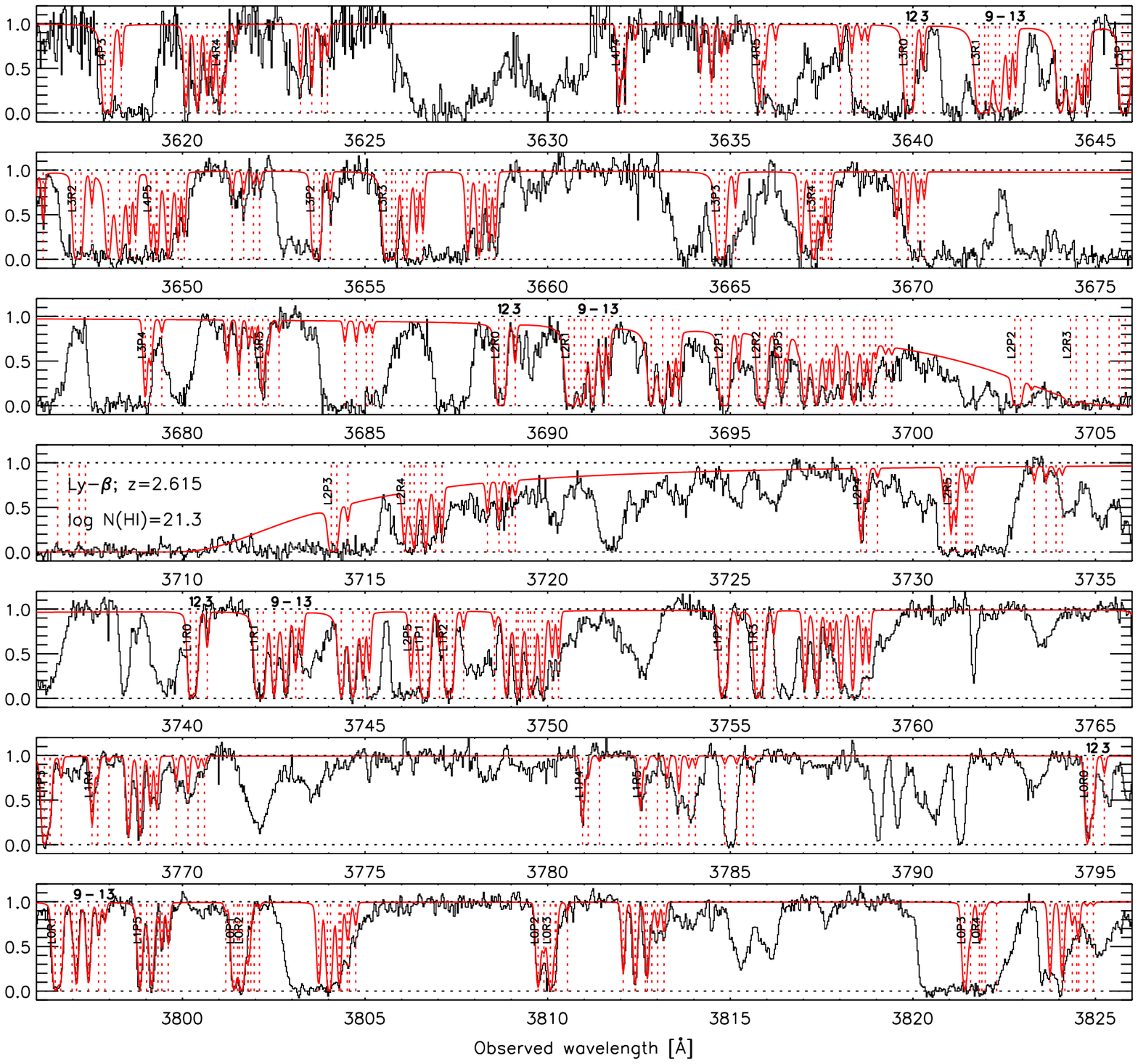}
 \caption{\label{q2348_H2} A portion of the normalised spectrum of
 Q\,2348$-$011. The resolving power is about $R\simeq$~47,000.  The H$_2$
spectrum derived from Voigt-profile fitting 
is superimposed to the quasar spectrum. The strong feature at ${\rm \lambda} \simeq 3707 {\rm \AA}$ 
is the \lyb\ absorption line from the DLA at $\zabs=2.6147$. 
H$_2$ components are numbered for the J~=~0 rotational level only. Transition
 names are abbreviated as LxYz standing for H$_2$\,Lx--0Yz, and only
 marked for component 1. The best fit parameters are given in Table~\ref{q2348H2t}.}
 \end{center}
\end{figure*}

The absorption profiles of the singly-ionised metal species have been fitted using fourteen
components and, in the following, we use their numbering to refer to
any individual velocity component. 
Seven of these components, spread over about 300~km\,s$^{-1}$, show associated H$_2$ absorption.
Multiple component Voigt-profile fittings have been performed
simultaneously on the H$_2$ absorption lines seen redward of the Lyman
break produced by the $\zabs = 2.6147$ system,
also including the Voigt profile of the \lyb\ line at $\zabs=2.6147$.
%
The continuum normalisation was performed using first-order broken
lines
between $\lambda_{\rm obs}\simeq$\,3600\,${\rm \AA}$ and $\lambda_{\rm
  obs}\simeq$\,3850\,${\rm \AA}$ and a routine based on the IRAF task {\it continuum} that can fit segments
of 50-100~\AA~ at once. 
For each $\sim$~50~${\rm \AA}$ wavelength range, the continuum is fitted by
approximately five to seven joint linear segments adjusted by an iterative procedure
that rejects absorptions over a certain threshold related to the local signal-to-noise ratio. 
This allows us to estimate the continuum
  position where the Lyman-$\alpha$ forest is dense, and over the
  $\zabs=2.6147$ Ly-$\beta$ absorption (see Fig.~\ref{q2348_H2}). Note that only the
H$_2$ Lyman band is covered by our spectrum.

\par The solution of a multi-component Voigt-profile fit is often not unique, especially
when absorption lines are blended and  sometimes saturated. In that case, it is 
difficult to constrain both the Doppler parameters and the column densities at the same 
time for all species. We therefore obtained a first guess of the molecular parameters
by fitting selected lines (see Table.~\ref{which_line}) of the first four rotational levels, J~=~0 to J~=~3, using Doppler 
parameters and redshifts derived from the \CI\ profiles, which are not saturated 
and located outside the Lyman-$\alpha$ forest. 
We then froze the corresponding parameters and added the J~=~4 and J~=~5 features. After a 
satisfactory fit to all H$_2$ transitions had been found in these conditions, we relaxed all 
parameters one by one, until the best fit was obtained. Then, a range of
column densities was derived by fitting the absorption
features with the extreme values of the Doppler parameter (see Table~\ref{q2348H2t}). 
The errors given in column 4 of Table~\ref{q2348H2t} therefore
correspond to a range of column densities and not to the RMS errors derived
from fitting the Voigt profiles. 
The quality of the overall fit can be assessed from Fig.~\ref{q2348_H2} and in Figs.~\ref{J0}
to \ref{J5}. Results are presented in Table~\ref{q2348H2t}.
The total H$_2$ column density integrated over all
rotational levels is $\log N$(H$_2$)~=~18.52$^{+0.29}_{-0.49}$, 
corresponding to a molecular fraction $\log f$~=~$-1.69^{+0.37}_{-0.58}$. 
We also derive a 3\,$\sigma$ upper limit on the column density
of HD molecules, leading to $\log N$(HD)/$N$(H$_2$) $<$ $-$3.3 for the
strongest H$_2$ components (\#\,1, 9 and 10).
We describe individual components below.

\subsubsection{Comments on individual components}
\emph{Components} 1 \& 2, $\zabs=2.42449(7)$ and
$2.42463(8)$: \label{C1} These two components 
are heavily blended and appear
as a single saturated feature for low H$_2$ rotational levels, but are better
separated for J~$\geq 4$, in \CI\ and low-ionisation metal lines (see Fig.~\ref{CI}, 
\ref{q2348metals} and \ref{J4}). Doppler parameters, $b$, are mainly constrained by the 
H$_2$\,L0--0\,R0, H$_2$\,L1--0\,P2,  H$_2$\,L1--0\,P3 profiles, and the transitions 
from the J~$\geq 4$ rotational levels, but they remain uncertain. No
damping wing is present. The first guess to fit the H$_2$ profiles, using redshifts from \CI,
\CI$^*$, and \CI$^{**}$ transitions, was extremely
helpful in this case. It must be emphasised that the blends of the two
components can be fitted using either low $b$ values, leading to high column 
densities or higher $b$ values, leading to lower column densities. 
The best fit is obtained, however, with $b$~=~2.9 km\,s$^{-1}$
and column densities $\log N$(H$_2$, J)~=~ 17.15; 17.70; 17.00;
17.60; 15.40 and 15.10 for J~=~0 to J~=~5, respectively. These
are the values we use in the following.
Rotational levels are unambiguously detected up to J~=~5. 
Possible absorptions from J~=~6 and J~=~7 levels, unfortunately,
are always redshifted on top of strong H~{\sc i} Lyman-$\alpha$ features
or in regions of low SNR. Upper limits are given in Table~\ref{q2348H2t}.\\
%
\emph{Component} 3, $\zabs=2.42491(7)$: This component is the weakest in the system and 
is not saturated. The Doppler parameter is therefore mainly constrained by the strongest
transition lines that are not blended: H$_2$\,L2--0\,R0, H$_2$\,L1--0\,R0,
and H$_2$\,L3--0\,P1. Only the first four rotational levels are
detected.
J~=~3 lines are always blended. However, the column-density can still
  be estimated using H$_2$\,L1--0\,R3 (left-half of the profile), H$_2$\,L1--0\,P3, and
  H$_2$\,L0--0\,R3 (see Fig.~\ref{J3}).\\
%
\emph{Components} 9 \& 10, $\zabs=2.42659(1)$ and $2.42688(5)$: These components 
have high H$_2$ column densities. They are separated well
and Doppler parameters are constrained well by the width
of the isolated and non-saturated absorption features H$_2$\,L0--0\,R0, H$_2$\,L0--0\,R1 and mostly of the P-branch ($\Delta J = -1$) Lyman-band transitions 
from the J~=~2 to J~=~4 rotational levels.
Note that component 9 of a particular transition is often blended with components 1 and 2 
of another transition from the next rotational level (see e.g., H$_2$\,L2--0\,R0 in Fig.~\ref{J0}).
The first five rotational levels (J~=~0 to J~=~4) are
detected. 
Some consistent features are seen for J~=~5 
(see transitions
H$_2$\,L4--0\,P5 and  H$_2$\,L2--0\,R5 on Fig.~\ref{J5}).
Upper-limits were derived as the largest value consistent
with the observed spectrum for levels J~=~5, J~=~6, and J~=~7.\\
\emph{Components} 12 \& 13, $\zabs=2.42713(5)$ and $2.42729(3)$: 
Absorptions from rotational levels J~=~0 to J~=~4 are
detected in the two components. $N$(J~=~2) is constrained mainly
  by H$_2$\,L0--0\,P2, but the maximum $N$-value is also constrained by H$_2$\,L3--0\,R2 and
  H$_2$\,L1--0\,P2, whereas the minimum $N$-value is constrained by the local depth of
  the features clearly seen in H$_2$\,L1--0\,R2 and H$_2$\,L0--0\,R2
  (see Fig.~\ref{J2}). Similar constraints are used for J~=~3. J~=~4 lines are blended or 
located in poor SNR
regions, but we can still constrain the column-density thanks to
H$_2$\,L4--0\,R4, H$_2$\,L4--0\,P4 and H$_2$\,L3--0\,P4. 
The J~=~5 rotational levels may be detected; but 
the corresponding column densities are very uncertain, and one should probably
consider these values only as upper limits (see Table~\ref{q2348H2t}).
\par A summary of the transitions used to constrain the
  column-density for each H$_2$-component in a
given rotational level is given in Table~\ref{which_line}.
The lines in this table are non-blended and 
have a rather good SNR, except when marked by ``u'' or ``l'' 
indicating that the line is used to estimate, respectively, an ``upper''
or a ``lower'' limit.

\begin{table}[!ht]
\caption{\label{which_line} Transitions used to constrain H$_2$ column
  densities in the seven components.
}
\begin{center}
\begin{tabular}{c l l l l l}
\hline \hline
Level   &  \#\,1 \& 2         &     \#\,3         &       \#\,9 \& 10         &    \#\,12        &    \#\,13      \\
\hline
J~=~0   & L0R0                &                   &  L0R0                     &  L0R0            & L0R0           \\ 
        & L1R0                & L1R0              &  L1R0                     &	 L1R0            &                \\
        & L2R0                & L2R0              &  L2R0                     &	 L2R0            & L2R0            \\
\hline
J~=~1   & L0P1                & L0P1              &                           &                  &                \\
        & L0R1		      & L0R1              &  L0R1                     &  L0R1            & L0R1           \\
        & L1P1		      & L1P1              &  L1P1		      &                  &                \\
        & L1R1		      &                   &  L1R1		      &  L1R1            &                \\
        & L2R1		      &                   &  L2R1                     &  L2R1            &            \\
        & L3P1		      & L3P1              &     		      &                  &                \\
        &       	      &                   &  L3R1		      &  L3R1            &                \\
\hline  
J~=~2   & L0P2                &                   & L0P2                      & L0P2             & L0P2           \\
        & L0R2                & L0R2              &                           & L0R2$^l$         & L0R2$^l$       \\
        & L1P2                & L1P2              & L1P2                      & L1P2	         & L1P2$^u$       \\
        & L1R2                &                   &                           & L1R2$^l$         & L1R2$^l$       \\
        & L3P2                & L3P2              &                           &     		 &                \\
        &                     &                   &                           & L3R2$^u$	 & L3R2$^u$       \\
\hline
J~=~3   &                     & L0R3              & L0R3                      & L0R3$^l$         & L0R3$^l$       \\ 
        & L1P3                & L1P3$^u$          & L1P3                      & L1P3             & L1P3$^u$       \\
        & L1R3                & L1R3              &                           &                  &                \\
        &                     &                   &                           & L3P3             & L3P3           \\
        & L3R3                &                   &                           & L3R3$^u$         &                \\
        &                     &                   & L4P3                      & L4P3             & L4P3           \\
\hline
J~=~4   & L1P4                & L1P4$^u$          &                           &                  &                \\
        & L1R4                & L1R4$^u$          &                           &			 &	          \\
        & L2P4                &                   &                           &			 &	          \\
        &                     &                   & L3P4                      & L3P4		 & L3P4$^u$       \\
        & L4P4                & L4P4$^u$          & L4P4                      & L4P4		 & L4P4           \\
        & L4R4                & L4R4$^u$          &                           & L4R4             & L4R4           \\
\hline
J~=~5   & L1R5                &                   &                           &                  &                 \\
        &                     &                   & L2R5$^u$                  & L2R5$^u$         &                 \\
        & L3R5                & L3R5$^u$          &                           &                  &                 \\
        &                     &                   & L4P5$^u$                  & L4P5$^u$         & L4P5$^u$        \\
\hline

\end{tabular}
\end{center}
\footnotesize
Transition names are abbreviated as LxYz standing for
H$_2$\,Lx--0\,Yz. \\
``u'' (resp. ``l'') means that the transition is used to give an upper
(resp. lower) limit on $N$(H$_2$; J).
\normalsize
\end{table} 

\begin{table}[!hc]
\caption{\label{q2348H2t} Voigt--profile fitting results for different
  rotational levels of the vibrational ground--state transition lines
  of H$_{2}$. }
\begin{center}
\begin{tabular}{ r l  c  l  c  c }
\hline
\hline

\# &$z_{\rm abs}$                &   Level   & $\log N$(H$_2$,~J)\,$^2$   & b               & $T_{0-J}$\,$^{3}$               \\
   &$v$\,$^{1}$ [km\,s$^{-1}$]     &           &                     &[km\,s$^{-1}$]   & [K]                     \\ 
\hline                                             			    						   
1& 2.42449(7) &   J = 0   &  15.55$-$17.40         &  2.9$^{+3.1}_{-0.4}$  & ...                    \\
 & $-$158     &   J = 1   &  16.40$-$18.00         &  ``             & 183$_{-145}$            \\
 &            &   J = 2   &  15.90$-$17.70         &  ``             & 262$_{-161}$    \\
 &            &   J = 3   &  16.10$-$17.80         &  ``             & 510$_{-340}$    \\
 &            &   J = 4   &  15.25$-$15.50         &  ``             & 274$^{+464}_{-35}$      \\
 &            &   J = 5   &  14.90$-$15.20         &  ``             & 311$^{+283}_{-35}$     \\
 &            &   J = 6   &  $<$14.00            &  ``             & $\leq$584              \\
 &            &   J = 7   &  $<$13.80            &  ``             & $\leq$610              \\
                                           			    						   
2& 2.42463(8) &   J = 0   &  15.35$^{+0.10}_{-0.10}$&  3.5$^{+1.0}_{-1.0}$   & ...                    \\
 &$-$145     &   J = 1   &  16.00$^{+1.50}_{-0.60}$&  ``             & 244$_{-170}$           \\
 &            &   J = 2   &  15.60$^{+0.10}_{-0.30}$&  ``             & 495$^{+398}_{-233}$    \\
 &            &   J = 3   &  15.30$^{+0.30}_{-0.10}$&  ``             & 324$^{+133}_{-41}$     \\
 &            &   J = 4   &  14.55$^{+0.10}_{-0.20}$&  ``             & 422$^{+54}_{-62}$      \\
 &            &   J = 5   &  14.30$^{+0.20}_{-0.20}$&  ``             & 433$^{+57}_{-45}$      \\
 &            &   J = 6   &  $<$13.80            &  ``             & $\leq$607              \\
 &            &   J = 7   &  $<$13.70            &  ``             & $\leq$648              \\
	       						    											   
3&2.42491(7)  &   J = 0   &  14.60$^{+0.10}_{-0.30}$&  1.3$^{+1.7}_{-0.4}$  & ...                    \\ 
 &$-$121     &   J = 1   &  15.00$^{+0.05}_{-0.50}$&  ``             & 134$^{+229}_{-69}$     \\ 
 &            &   J = 2   &  14.30$^{+0.10}_{-0.20}$&  ``             & 223$^{+149}_{-51}$     \\ 
 &            &   J = 3   &  14.55$^{+0.10}_{-0.25}$&  ``             & 324$^{+133}_{-66}$      \\ 
 &            &   J = 4   &  $\leq$13.80            &  ``             & $\leq$509              \\ 
 &            &   J = 5   &  $\leq$13.70            &  ``             & $\leq$525              \\ 
                                           			    						   
9&2.42659(1)  &   J = 0   &  17.30$^{+0.20}_{-0.60}$&  2.1$^{+0.9}_{-1.0}$  & ...                    \\ 
 &$+$25      &   J = 1   &  17.70$^{+0.10}_{-0.30}$&  ``             & 134$_{-63}$     \\ 
 &            &   J = 2   &  16.50$^{+0.90}_{-0.50}$&  ``             & 148$_{-47}$    \\ 
 &            &   J = 3   &  16.70$^{+0.75}_{-0.50}$&  ``             & 231$^{+546}_{-62}$     \\ 
 &            &   J = 4   &  14.55$^{+0.30}_{-0.25}$&  ``             & 200$^{+64}_{-22}$      \\ 
 &            &   J = 5   &  $\leq$ 14.00      &  ``             & $\leq$264               \\ 
 &            &   J = 6   &  $<$13.50            &  ``             & $\leq$361              \\ 
 &            &   J = 7   &  $<$13.70            &  ``             & $\leq$446              \\ 
                                          			     					   
10&2.42688(5) &   J = 0   &  17.30$^{+0.20}_{-0.60}$&  1.9$^{+1.1}_{-0.4}$    & ...                    \\ 
  &$+$51     &   J = 1   &  17.90$^{+0.20}_{-0.20}$&  ``             & 209$_{-111}$      \\ 
  &           &   J = 2   &  17.40$^{+0.20}_{-0.40}$&  ``             & 371$_{-186}$           \\ 
  &           &   J = 3   &  17.00$^{+0.30}_{-0.50}$&  ``             & 274$^{+341}_{-83}$     \\ 
  &           &   J = 4   &  14.85$^{+0.30}_{-0.35}$&  ``             & 218$^{+78}_{-30}$      \\ 
  &           &   J = 5   &  $\leq$ 14.00       &  ``             &  $\leq$264            \\ 
  &           &   J = 6   &  $<$13.50            &  ``             & $\leq$361              \\ 
  &           &   J = 7   &  $<$13.70            &  ``             & $\leq$446              \\ 
                                           			     					   
12&2.42713(5) &   J = 0   &  14.95$^{+0.10}_{-0.05}$&  2.5$^{+1.0}_{-1.0}$    & ...                    \\ 
  &$+$73     &   J = 1   &  15.35$^{+0.05}_{-0.10}$&  ``             & 134$^{+29}_{-35}$     \\ 
  &           &   J = 2   &  14.90$^{+0.10}_{-0.30}$&  ``             & 297$^{+74}_{-103}$     \\ 
  &           &   J = 3   &  15.10$^{+0.10}_{-0.20}$&  ``             & 379$^{+56}_{-77}$    \\ 
  &           &   J = 4   &  14.20$^{+0.10}_{-0.30}$&  ``             & 435$^{+42}_{-83}$      \\ 
  &           &   J = 5   &  $\leq$13.70             &  ``             & $\leq$408              \\ 

13&2.42729(3) &   J = 0   &  14.65$^{+0.05}_{-0.10}$&  2.0$^{+1.0}_{-1.0}$  & ...                    \\ 
  &$+$87     &   J = 1   &  15.35$^{+0.10}_{-0.30}$&  ``             & 291$^{+1075}_{-169}$           \\ 
  &           &   J = 2   &  15.00$^{+0.10}_{-0.35}$&  ``             & 637$^{+855}_{-340}$           \\ 
  &           &   J = 3   &  15.20$^{+0.10}_{-0.30}$&  ``             & 576$^{+201}_{-180}$    \\ 
  &           &   J = 4   &  14.10$^{+0.10}_{-0.30}$&  ``             & 493$^{+76}_{-93}$              \\ 
  &           &   J = 5   &  $\leq$13.70             &  ``             & $\leq$469              \\ 
\hline
\end{tabular}
\end{center}
$^1$ Velocity relative to $\zabs=2.42630$\\
$^2$ Best fit values with errors representing the allowed range on
$\log N$(H$_2$), not the rms error on the fit. Due to the large uncertainties, only a range is given for
component 1 (see text).\\
$^3$ Temperatures corresponding to the best-fit values. In some cases,
the upper-limit on the ratio $N$(H$_2$, J)$/N$(H$_2$,
  J\,=\,0) is higher than the maximum expected ratio in case of local
  thermal equilibrium (see Eq.~\ref{texeq}), and the corresponding
  upper-limit on $T_{\rm 0-J}$ doesn't exist. 

\normalsize
\end{table}

\subsection{Neutral carbon}
\begin{figure}[!ht]
 \begin{center}
\includegraphics[width=\hsize,clip=]{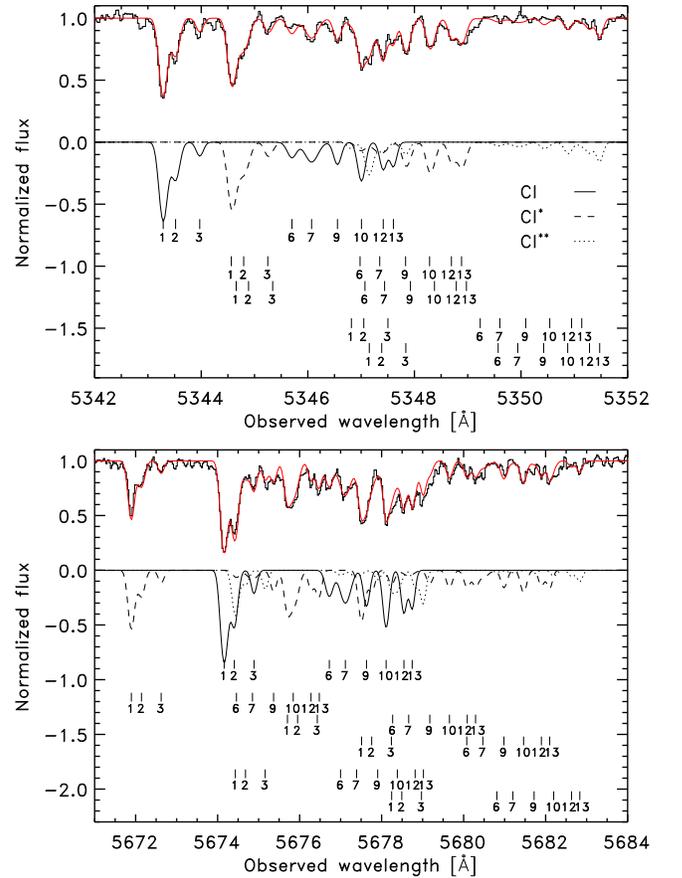}
\caption{\label{CI} Observed portions of
  the normalised spectrum of Q\,2348$-$011 (upper part in both panels) 
where \CI\ absorptions are seen, with the model spectrum superimposed. The 
modelled spectrum is decomposed into \CI\ (solid line), \CI$^*$ (dashed line), and
\CI$^{**}$ (dotted line) model spectra (lower part in both
  panels). The positions of the nine components are marked by small vertical lines. The associated labels 
correspond to the numbering in Table~\ref{q2348CIt} (see also Tables~\ref{q2348H2t} and
  \ref{q2348metalst}).}
 \end{center}
\end{figure}

Neutral carbon is detected in nine components. The corresponding absorption profiles are 
complex, as resulting from the blend of absorption lines from three fine-structure levels of the 
neutral carbon ground state (2s$^2$2p$^2$\,$^3$P$_{0,1,2}$).
In Fig.~\ref{CI} we show the fit to the
\CI\ absorption features around $\lambda_{\rm obs} \approx$~5346~${\rm
  \AA}$ and $\lambda_{\rm obs} \approx$~5676~${\rm \AA}$ 
with contributions from the eleven transitions: \CI$\lambda$1560, \CI$\lambda$1656,
\CI$^{*}\lambda\lambda$1560.6,1560.7,
\CI$^{*}\lambda\lambda\lambda$1656,1657.3,1657.9,
\CI$^{**}\lambda\lambda$1561.3,1561.4, and \CI$^{**}\lambda\lambda$1657,1658. 
We first 
performed a fit with seven \CI\ components corresponding to the seven H$_2$
detected components, but we left the exact positions of the \CI\
components free.
We then added \CI$^{*}$, and finally \CI$^{**}$ transitions. 
To reproduce the overall profile well, we had to introduce two weak \CI\ 
components (\# 6 and 7) that are not detected in H$_2$, but are
detected in the absorption profiles of low-ionisation metals. We then performed a second fit with nine components, in the same order: first
\CI\ only, then adding \CI$^*$ and finally \CI$^{**}$.
Finally all transitions were fitted together. The resulting small velocity
shifts between \CI\ and H$_2$ components (see Table~\ref{q2348CIt}) 
are within uncertainties. 
This means that 
H$_2$ and \CI\ absorptions are generally produced in the same regions. 
Note, however, that the Doppler parameters found for the \CI\ components are
slightly larger than those found for the H$_2$ components, indicating
that the \CI\ regions are probably of slightly larger dimensions.
Results of the fit (Fig.~\ref{CI}) are presented in Table~\ref{q2348CIt}.
Three $\sigma$ upper-limits are derived for $N$(\CI$^{**}$) in
components 6, 7, and 9, leading to upper-limits on $N$(\CI$^{**}$)/$N$(\CI) (see Table~\ref{q2348CIt}).

\begin{table*}[!ht]
\caption{\label{q2348CIt} Column densities of the \CI\ fine-structure levels.
} 
\begin{center}
\begin{tabular}{r c  c  c  c  c  c  c  c}
\hline
\hline

\# & $\zabs$  & b  &  $\log N$(\CI) &$\log N$(\CI$^*$) & $\log N$(\CI$^{**}$)\,$^1$ & $\Delta v_{\ion{C}{i}/{\rm H}_2}\,^{2,3}$  & $\log N$(\CI$^*$)/N(\CI) & $\log N$(\CI$^{**}$)/N(\CI)    \\
   &          & [km\,s$^{-1}$]&     &                  &                      & [km\,s$^{-1}$]                     &                          &                                \\
\hline

1  & 2.42450(5) & 4.2$\pm$ 0.3    &     13.55$\pm$ 0.02   &   13.51$\pm$ 0.03   &    13.10$\pm$ 0.04     &       +0.7 &   -0.04$\pm$0.04   & -0.45$\pm$0.04  \\
2  & 2.42465(4) & 4.6$\pm$ 0.4    &     13.10$\pm$ 0.06   &   13.06$\pm$ 0.02   &    12.54$\pm$ 0.10     &       +1.4 &   -0.04$\pm$0.06   & -0.56$\pm$0.12  \\
3  & 2.42494(4) & 2.8$\pm$ 0.4    &     12.55$\pm$ 0.08   &   12.66$\pm$ 0.09   &    12.60$\pm$ 0.12     &       +2.4 &    0.11$\pm$0.12   &  0.05$\pm$0.14  \\
6  & 2.42605(3) & 5.9$\pm$ 0.5    &     12.70$\pm$ 0.07   &   12.50$\pm$ 0.09   &    $<$12.50            &        --- &   -0.20$\pm$0.11   & $<$-0.20        \\
7  & 2.42628(9) & 8.4$\pm$ 0.5    &     12.96$\pm$ 0.04   &   12.75$\pm$ 0.04   &    $<$12.50            &        --- &   -0.21$\pm$0.06   & $<$-0.46        \\
9  & 2.42660(0) & 4.0$\pm$ 0.5    &     12.80$\pm$ 0.02   &   12.93$\pm$ 0.10   &    $<$12.50            &       +0.8 &    0.13$\pm$0.10   & $<$-0.30        \\
10 & 2.42688(9) & 4.0$\pm$ 0.5    &     13.09$\pm$ 0.02   &   13.05$\pm$ 0.04   &    12.55$\pm$ 0.20     &       +0.3 &   -0.04$\pm$0.04   & -0.54$\pm$0.20  \\
12 & 2.42715(1) & 3.2$\pm$ 0.5    &     12.87$\pm$ 0.04   &   12.84$\pm$ 0.04   &    12.53$\pm$ 0.05     &       +1.4 &   -0.03$\pm$0.06   & -0.34$\pm$0.06  \\
13 & 2.42727(3) & 3.0$\pm$ 0.6    &     12.80$\pm$ 0.05   &   12.90$\pm$ 0.03   &    12.80$\pm$ 0.10     &       -1.7 &    0.10$\pm$0.06   &  0.00$\pm$0.11  \\

\hline
\end{tabular}
\end{center}
\footnotesize
$^1$ 3\,$\sigma$ upper-limits are derived for $N$(\CI$^{**}$) in components 6, 7 and 9.\\
$^2$ We estimate the errors on $\Delta v_{\ion{C}{i}/{\rm H}_2}$ to be
 about 2 to 3 km\,s$^{-1}$.\\
$^3$ There is no velocity shift between the
  components of neutral carbon and those of molecular
  hydrogen.\\
\normalsize
\end{table*}

\subsection{Metal content}
\begin{figure}[!ht]
\begin{center}
\includegraphics[width=\hsize,clip=]{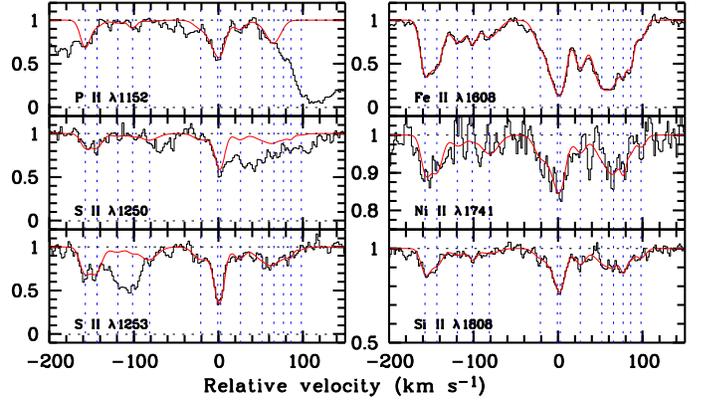}
\caption{\label{q2348metals} Low-ionisation metal transitions in the $\zabs=2.43$ DLA 
toward Q\,2348$-$011. 
Note that the strongest
    component in the metal profiles (at $v=0$~km\,s$^{-1}$) does not show associated
H$_2$ absorption (see Fig.~\ref{q2348CIVSiIV}). We give upper-limits
on $N$(\PII) for the three
    redder components of \PII\ as they are blended with intervening H~{\sc i} 
    absorptions. \SII\ transitions are also located in the \lya\ forest;
    but with the help of the two transitions shown above,
    we can still derive column densities or upper-limits for each component. Corresponding Voigt-profile parameters 
are given in Table~\ref{q2348metalst}.}
\end{center}
\end{figure}

The absorption profiles of low-ionisation species (\PII,
\SII, \FeII, \NiII\ and \SiII) are
complex and span about $\Delta v \simeq$~300 km\,s$^{-1}$ in velocity space.
Fourteen components were needed to reconstruct the whole profiles,
i.e., the nine \CI\ components plus five components that are not
detected in \CI. 
A few profiles are shown in Fig.~\ref{q2348metals} and results from the fit are given in Table~\ref{q2348metalst}.
The overall metallicity is found to be $-$0.80, $-$0.62,
and $-$1.17$\pm$0.10 for, respectively, [Si/H], [S/H], and [Fe/H].
The strongest component (\#\,7) in these profiles does not correspond to any H$_2$ 
component and is associated with a weak \CI\ component.
Redshifts have been relaxed to allow for the best fit to be obtained.
Shifts between singly ionised metals components and \CI\ components are 
well within errors, however.
Voigt-profile parameters given in Table~\ref{q2348metalst} are well-constrained for species with transitions located outside the \lya\ forest,
i.e. \FeII, \NiII, and \SiII. The \SII\ and \PII\ transitions are partly blended with
intervening \lya\ \HI\ absorptions. Nevertheless, all \SII\ components
can be constrained thanks to the observations of two different transitions. The three redder components of \PII\
are blended, and we cannot derive $N$(\PII) for these
components (see Table~\ref{q2348metalst}).

\begin{figure}[!ht]
\begin{center}
\begin{tabular}{c}
\includegraphics[width=0.6\hsize,clip=]{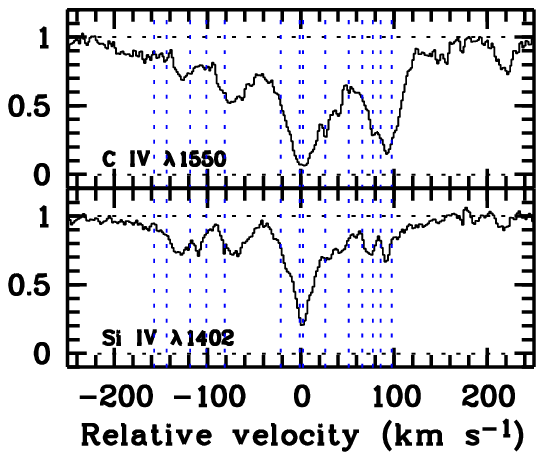}\\
\includegraphics[width=0.6\hsize,clip=]{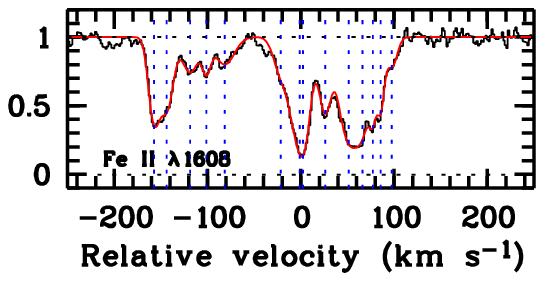}\\
\includegraphics[width=0.6\hsize,clip=]{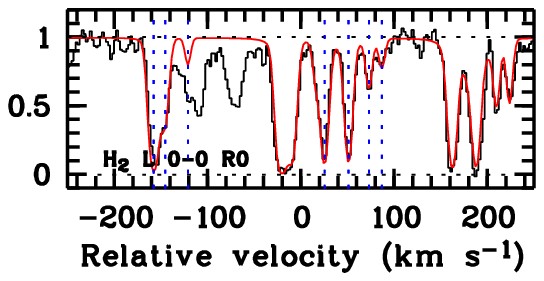}\\
\end{tabular}

\caption{\label{q2348CIVSiIV} Absorption profiles of \SiIV\ and \CIV\
  in the $\zabs=2.43$ DLA toward Q\,2348$-$011. The vertical lines
  mark the positions of low-ionisation metal components in the
  \CIV\, \SiIV\ and \FeII\ panels. They mark the positions of the seven H$_2$
  components in the H$_2$\,L0--R0 panel.
}

\end{center}
\end{figure}

Figure~\ref{q2348CIVSiIV} shows the overall absorption profile of highly
ionised species: \CIV\ and \SiIV. It can be seen that the \CIV\ counterpart of 
H$_2$ components are generally weak compared to the other components.
The strongest \CIV\ feature at $v$~=~0~km\,s$^{-1}$ does not correspond to any H$_2$ component
but corresponds to the strongest component in Fe~{\sc ii}.


\begin{table}[!ht]
\caption{\label{q2348metalst} Results of the Voigt-profile fitting of
  the singly-ionised metal lines.}
\scriptsize
\begin{center}
\begin{tabular}{r l l l  c c }
\hline
\hline
\#&$z_{abs}$               & Ion (X)  & $\log N(X)$           &  b  &
$\Delta v_{\rm X/C\,{\sc I}}$ \\
  &$v$~$^1$[km\,s$^{-1}$]&             &                       & [km\,s$^{-1}$] & [km\,s$^{-1}$]\\ 
\hline
1&2.42450(5) &       \PII        & 12.84$\pm$ 0.03 &   6.6$\pm$0.1       & $+$0.0\\
 &  $-$157         &       \SII        & 14.11$\pm$ 0.02 &   ``                &     \\
 &                 &       \FeII       & 13.73$\pm$ 0.01 &   ``                &     \\
 &                 &       \NiII       & 12.86$\pm$ 0.05 &   ``                &     \\
 &                 &       \SiII       & 14.30$\pm$ 0.03 &   ``                &     \\
		     		    
2&2.42465(7) &       \PII        & 12.22$\pm$ 0.09 &   8.1$\pm$0.2       & $+$0.3\\
 & $-$144          &       \SII        & 14.21$\pm$ 0.02 &   ``                &     \\
 &                 &       \FeII       & 13.70$\pm$ 0.01 &   ``                &     \\
 &                 &       \NiII       & 12.92$\pm$ 0.04 &   ``                &     \\
 &                 &       \SiII       & 14.18$\pm$ 0.04 &   ``                &     \\
		     		    
3&2.42494(4) &       \PII        & 11.97$\pm$ 0.18 &   10.7$\pm$0.7      & $+$0.0\\
 & $-$119          &       \SII        & $<$ 13.60    &   ``                &     \\
 &                 &       \FeII       & 13.39$\pm$ 0.01 &   ``                &     \\
 &                 &       \NiII       & $<$12.50     &   ``                &     \\
 &                 &       \SiII       & 13.87$\pm$ 0.08 &   ``                &     \\
		     		    
4&2.42514(4)       &       \PII        & 12.15$\pm$ 0.09 &   5.1$\pm$0.1       & --- \\
 & $-$101          &       \SII        & $<$ 13.60    &   ``                &     \\
 &                 &       \FeII       & 13.11$\pm$ 0.01 &   ``                &     \\
 &                 &       \NiII       & $<$12.50     &   ``                &     \\
 &                 &       \SiII       & 13.82$\pm$ 0.07 &   ``                &     \\
		     		    
5&2.42536(8)       &       \PII        & 11.68$\pm$ 0.48 &   13.5$\pm$0.2      & --- \\
 & $-$82           &       \SII        & $\leq$14.04      &   ``                &     \\
 &                 &       \FeII       & 13.35$\pm$ 0.01 &   ``                &     \\
 &                 &       \NiII       & 12.81$\pm$ 0.07 &   ``                &     \\
 &                 &       \SiII       & 13.85$\pm$ 0.09 &   ``                &     \\
                                         
6&2.42605(6)       &       \PII        & 12.32$\pm$ 0.09 &   9.8$\pm$0.2       & $+$0.3\\
 & $-$21           &       \SII        & 13.85$\pm$ 0.05 &   ``                &     \\
 &                 &       \FeII       & 13.34$\pm$ 0.01 &   ``                &     \\
 &                 &       \NiII       & 12.71$\pm$ 0.08 &   ``                &     \\
 &                 &       \SiII       & 13.93$\pm$ 0.07 &   ``                &     \\
                               
7&2.42628(8)       &       \PII        & 13.09$\pm$ 0.04 &   11.9$\pm$0.3     & $-$0.1\\
 &  $-1$           &       \SII        & 14.08$\pm$ 0.07 &   ``               &     \\
 &                 &       \FeII       & 14.10$\pm$ 0.01 &   ``               &     \\
 &                 &       \NiII       & 13.18$\pm$ 0.06 &   ``               &     \\
 &                 &       \SiII       & 14.55$\pm$ 0.04 &   ``               &     \\
		     		    
8&2.42632(7)       &       \PII        & 12.45$\pm$ 0.12 &   5.7$\pm$0.1      & --- \\
 & $+$2            &       \SII        & 14.49$\pm$ 0.03 &   ``               &     \\
 &                 &       \FeII       & 13.58$\pm$ 0.01 &   ``               &     \\
 &                 &       \NiII       & $<$12.50     &   ``               &     \\
 &                 &       \SiII       & 14.14$\pm$ 0.08 &   ``               &     \\
          
                                
9&2.42660(0)       &       \PII        & 12.35$\pm$ 0.06 &   7.4$\pm$1.0      & $+$0.0\\
 & $+$26           &       \SII        & 13.78$\pm$ 0.06 &   ``               &     \\
 &                 &       \FeII       & 13.68$\pm$ 0.01 &   ``               &     \\
 &                 &       \NiII       & 12.55$\pm$ 0.09 &   ``               &     \\
 &                 &       \SiII       & 14.12$\pm$ 0.04 &   ``               &     \\

10&2.42688(8)      &       \PII        & 12.00$\pm$ 0.03 &   12.0$\pm$2.2     & $-$0.1\\
  &  $+$51         &       \SII        & 13.90$\pm$ 0.05 &   ``               &     \\
  &                &       \FeII       & 14.09$\pm$ 0.05 &   ``               &     \\
  &                &       \NiII       & 12.81$\pm$ 0.07 &   ``               &     \\
  &                &       \SiII       & 14.20$\pm$ 0.15 &   ``               &     \\

11&2.42705(4)      &       \PII        & 12.90$\pm$ 0.05 &   10.5$\pm$3.5     & --- \\
  &$+$66           &       \SII        & 14.04$\pm$ 0.05 &   ``               &     \\
  &                &       \FeII       & 13.92$\pm$ 0.04 &   ``               &     \\
  &                &       \NiII       & 12.97$\pm$ 0.05 &   ``               &     \\
  &                &       \SiII       & 14.23$\pm$ 0.05 &   ``               &     \\

12&2.42718(3)      &       \PII        & blend           &   2.8$\pm$1.5      & $+$2.8\\
  &$+$77           &       \SII        & $<$13.60     &       ``           &     \\
  &                &       \FeII       & 13.38$\pm$ 0.01 &   ``               &     \\
  &                &       \NiII       & 12.53$\pm$ 0.08 &   ``               &     \\
  &                &       \SiII       & 13.97$\pm$ 0.04 &   ``               &     \\
		     		    
13&2.42727(9)      &       \PII        & blend           &3.4$\pm$1.3      & $+$0.5\\
  & $+$86          &       \SII        & $<$13.60     &    ``              &     \\
  &                &       \FeII       & 13.47$\pm$ 0.01 &   ``               &     \\
  &                &       \NiII       & $<$ 12.50    &   ``               &     \\
  &                &       \SiII       & 13.68$\pm$ 0.08 &   ``               &     \\
		     		    
14&2.42741(5)      &       \PII        & blend           &   6.4$\pm$1.2      & --- \\
  & $+$98          &       \SII        & $<$13.60     &   ``               &     \\
  &                &       \FeII       & 13.14$\pm$ 0.01 &   ``               &     \\
  &                &       \NiII       & $<$ 12.50    &   ``               &     \\
  &                &       \SiII       & 13.86 $\pm$0.06 &   ``               &     \\
\hline
\end{tabular}
\normalsize
\end{center}
$^1$ \footnotesize Relative velocity (in km~s$^{-1}$) to the center
  of the Lyman-$\alpha$ absorption at $\zabs=2.42630$.\\
Upper-limits are given at 3\,$\sigma$, except in the case of \SII\ in
  component 5, for which the column-density derived has been
  considered as an upper-limit to take account of a possible blend.
\end{table}

\begin{figure}[!ht]
\begin{center}
\includegraphics[width=\hsize,clip=]{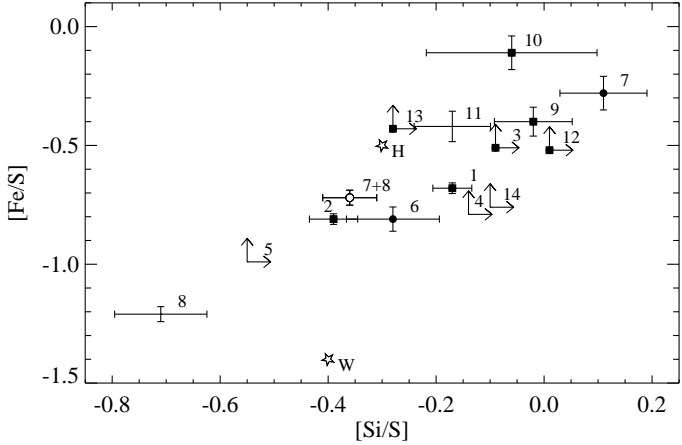}
\caption{\label{depletion} The iron-to-sulfur abundance ratio, [Fe/S],
is plotted as a function of the silicon-to-sulfur abundance ratio, [Si/S].
The observed correlation is interpreted as the result of depletion of 
Fe and Si onto dust grains.
Different velocity components are numbered from 1 to 14. Filled
  squares are for components where both H$_2$ and \CI\ are detected, while
  filled circles are for components where \CI\ is detected but not
  H$_2$. The open circle stands for components 7 and 8 considered
  as a single velocity component.
  Stars mark the typical depletions observed in the Galactic 
halo clouds (H) and in clouds of the Galactic warm disk \citep[W; see Table~6 of][]{Welty99}. The point corresponding to the depletion 
in clouds of the Galactic cold disk would be located off of the graph with [Si/S$]$~=~$-1.3$ and [Fe/S$]$~=~$-2.2$.}

\end{center}
\end{figure}

The iron-to-sulfur abundance ratio is plotted in Fig.~\ref{depletion} as a function of the 
silicon-to-sulfur abundance ratio. The abundance ratios are defined as
[X/S]$=\log N($X$)/N($S$)_{\rm obs}-\log  N($X$)/N($S$)_{\odot}$.
A strong correlation between [Fe/S] and [Si/S] is observed, similar
to what has been observed previously by \citet{Petitjean02} and \citet{Rodriguez06}. 
This correlation is easily interpreted as a depletion pattern of Fe and Si onto dust grains. 
Note 
that H$_2$ is detected in components were iron is not strongly depleted with
[Fe/S] as high as $-$0.1 in component \#\,10 and that the highest 
depletions are found in
  components where H$_2$ is not detected, components \#\,5 and 8. 
However, components \#\,7 and 
\#\,8 are heavily blended and the distribution of $N$(\SII) between these two
  components is uncertain. If we consider these two components as a
  single feature, then the corresponding depletion would be lower than for component
\#\,8. 
In component \#\,5, the \SII\ column-density
  might be overestimated due to poor SNR and a small blend (see
  Fig.~\ref{q2348metals}). 
In addition, it can be seen in Fig.~3 that component
\#\,10 is blended in the wing of component \#\,11. Therefore the decomposition 
is difficult, and the $N$(Fe~{\sc ii}) column density in this component could 
be overestimated. 
\par
The typical abundance patterns in the Galactic halo, warm disk clouds, and cold
disk clouds are also indicated in Fig.~\ref{depletion}. As for Magellanic clouds, 
the halo pattern represents the depletion in individual DLA
components best. 

\subsection{Chlorine}
When H$_2$ is optically thick, singly-ionised chlorine (\ClII) 
will go through rapid charge exchange reaction with H$_2$ to produce
\HI\ and \ClI \citep[see][]{Jura74a}. Neutral chlorine should therefore be a good tracer
of H$_2$ \citep{Jura78} and provide additional information for
constraining the physical parameters of the gas.
Unfortunately, the wavelength at which the strongest chlorine line
(\ClI\,$\lambda$1347) is expected to fall is not covered by our spectrum.
It is possible to give an upper-limit on $N$(\ClI), thanks to \ClI$\lambda$1088 and
\ClI$\lambda$1101 transitions whose oscillator strengths have recently been
measured by \citet{Sonnentrucker06}. We derive $N$(\ClI)~$\leq 10^{13}$~cm$^{-2}$ for each 
component at the 3\,$\sigma$ detection limit. This leads to
about $N$(\ClI)$/N$(H$_2$)~$\leq  10^{-5}$ for the strongest H$_2$
components (1, 9, and 10). Typical ratios measured along lines of
sight toward the Magellanic clouds and the Galactic disk are, respectively,
$N$(\ClI)$/N$(H$_2$)~$\sim 0.4-3 \times 10^{-6}$ \citep{Andre04} and
$N$(\ClI)$/N$(H$_2$)~$\sim 0.4\times10^{-6}$ \citep{Sonnentrucker02}. 
The non-detection of \ClI\ with such an upper-limit is therefore not surprising.  
Singly-ionised chlorine (\ClII), in turn, has only two transitions at $\lambda_{\rm rest}\simeq$~1063~${\rm \AA}$ and
1071~${\rm \AA}$. The corresponding features are blended with H$_2$ or \lya\ lines, but
an upper-limit at 3\,$\sigma$ of about $N$(\ClII)~$\leq 10^{14}$~cm$^{-2}$ can still be derived for
components with a relative velocity between $v \simeq$ $-120$~km\,s$^{-1}$ and $-30$~km\,s$^{-1}$
(i.e. components 3, 4, and 5).



\section{Determination of physical parameters}

\subsection{Excitation temperatures}
\label{parH2}

The excitation temperature $T_{XY}$ between two rotational levels $X$ and $Y$
is defined as
\begin{center}
\begin{equation}
{N_Y\over N_X}= {g_Y\over g_X} e^{-E_{XY}/{kT_{XY}}}
\label{texeq}
\end{equation}
\end{center}

\begin{figure}[!ht]
\begin{center}
\includegraphics[width=\hsize,clip=]{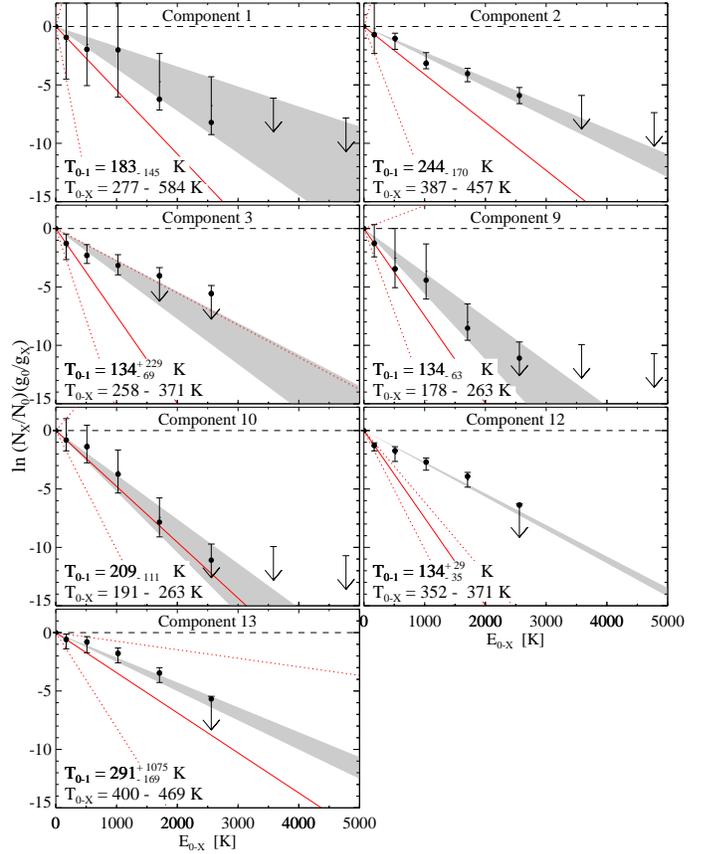}
\caption{\label{texfig} Excitation diagrams for the seven H$_2$ components.
The grey area shows the range of excitation temperatures
that can explain the populations of the J~$\geq$~2 rotational levels. 
The slope of straight lines give directly $1/{\rm T_{0-X}}$. 
The solid line stands for
$T_{0-1}$ and dashed lines for the upper and lower limits on
$T_{0-1}$. 
}
\end{center}
\end{figure}

\par \noindent where $g_{X}$ is the statistical weight of rotational level $X$:
$g_X=(2$J$_X+1)(2$I$_X+1)$ with nuclear spin I~=~0 for even J (para-H$_2$)
and I~=~1 for odd
J (ortho-H$_2$), $k$ is the Boltzman constant, and $E_{XY}$ is the
energy difference between the levels J~=~$X$ and J~=~$Y$.
If the excitation processes are dominated by collisions, 
then the populations of the rotational levels follow a
Boltzman distribution described by a unique excitation temperature for
all rotational levels. 
In Fig.~\ref{texfig}, we plot, for each H$_2$ component, the natural logarithm of
the ratios $N($H$_2,$~J~$=X)/N($H$_2, $~J~$=0)$ weighted by the degeneracy
factor $g_X/g_0$ against the energy between rotational levels. The
slope of a straight line starting at the origin gives the inverse of the excitation temperatures $T_{0-X}$ directly. 

Figure~\ref{texfig} and Table~\ref{q2348H2t} show that $T_{0-1}$ is
generally much lower than the other excitation
temperatures. Generally, $T_{0-1}$ is a good indicator of the kinetic temperature
\citep{Roy06, Lepetit06}. However, two effects could act, in opposite
directions, against this simple assumption.
If the J~=~1 level is not thermalised, then $T_{01}$ is
  a lower-limit on $T_{\rm kin}$. On the contrary, the selective 
self-shielding of molecular hydrogen
 in the J~=~1 level for column densities below $N$(H$_2)\approx 10^{16}$~cm$^{-2}$ 
should make $T_{01}$ an upper-limit on $T_{\rm kin}$.
Given the large densities we derive in those components (see \S\,\ref{parCI}),
collisions are not negligible and contribute towards thermalising 
the two levels.
Despite large allowed ranges for individual components, and a
  relatively large dispersion among the individual components, the ratios
$N$(H$_2$, J~=~1)$/N$(H$_2$, J~=~0) for the seven H$_2$ components are
all consistent with $T_{01} \approx 140$~K, which corresponds to the intersection of all the individual
  $T_{01}$ ranges.
This value can be taken as representative of the kinetic
  temperature for the derivation of the densities (see \S\,\ref{parCI}).
The kinetic temperatures, roughly estimated by $T_{01}$ (see
  Table~\ref{q2348H2t}), are nevertheless similar to what was found in previous studies.
\citet{Srianand05}, using the ortho-to-para ratio for thirteen H$_2$
components in seven DLAs, found $T\approx 153\pm78$~K
\citep[see also][$T \sim 90-180$~K]{Ledoux06b}. These values are higher than
those measured in the ISM \citep[77$\pm$17~K;][]{Rachford02} and in the Magellanic Clouds
\citep[82$\pm$21~K;][]{Tumlinson02}. On the other hand, they are similar 
to what is observed through high-latitude Galactic sight lines
  (124$\pm$8~K; \citet{Gillmon06} or ranging from 81~K at $\log
  N$(H$_2$)~=~20 to 219~K at $\log N$(H$_2$)~=~14; \citet{Wakker06}).
This could be due to a high UV radiation field but also to different grain properties. 
Indeed, the photo-electric
  heating of the medium should be more efficient in the case of smaller
  grains. 
The higher excitation temperatures measured for higher rotational
levels (J~$\geq$~2) show that other processes than collisions are in play. 
Rotational levels J~=~4 and J~=~5 can be populated by cascades
following UV or formation pumping. We indeed assume in the
following that these are the main processes populating high-J levels.
It is nevertheless also possible that the H$_2$ rotational excitation
is due to C-shocks or turbulence \citep{Joulain98,Cecchi-Pestellini05}
as in diffuse clouds.


\subsection{Photo-absorption rate}

We can estimate the total absorption rate of Lyman- and Werner-band UV photons
by H$_2$ in the J~=~0 and J~=~1 levels, $\beta_0$ and $\beta_1$, respectively, following 
\citet{Jura75}. The equilibrium of the J~=~4 and 5 level populations is described by

\begin{equation}
p_{4,0}\,\beta_0\,n({\rm H}_2, {\rm J=0})+0.19\,R\,n({\rm H})\,n = A_{4\rightarrow2}\,n({\rm H}_2, {\rm J=4})
\label{beta04eq}
\end{equation}

and

\begin{equation}
p_{5,1}\,\beta_1\,n({\rm H}_2, {\rm J=1})+0.44\,R\,n({\rm H})\,n = A_{5\rightarrow3}\,n({\rm H}_2, {\rm J=5})
\label{beta15eq}
\end{equation}

\par \noindent where $R$ is the H$_2$ molecule formation rate, $n$ the proton density, $p$ the pumping coefficients 
and $A$ the spontaneous transition probabilities. 

The factor 0.19 (resp. 0.44) means that about 19\% (resp. 44\%) of
H$_2$ molecules form in the J~=~4 (resp. J~=~5) level \citep{Jura74b}.
We use the spontaneous transition probabilities
$A_{4\rightarrow2} = 2.8 \times 10^{-9}$\,s$^{-1}$ and 
$A_{5\rightarrow3} = 9.9 \times 10^{-9}$\,s$^{-1}$ \citep{Spitzer78} and the
pumping efficiencies 
$p_{4,0}=0.26$ and $p_{5,1}=0.12$ \citep{Jura75}.
The pumping due to formation of molecules onto dust grains is often neglected if densities
are lower than $5 \times 10^3$\,cm$^{-3}$, which is 
generally the case in 
\dla\ systems \citep{Srianand05}. 
However, in the following, we rewrite the above equations 
in a form that only depends on the relative populations of H$_2$ rotational levels.

The equilibrium between formation and destruction of H$_2$ can be written
\begin{equation}
 R\,n\,n({\rm H}) = R_{\rm diss}\,n({\rm H}_2)
\label{eq_fd_h2}
\end{equation}

\par \noindent where $R_{\rm diss}=0.11\,\beta_0 $, assuming that 11\% of the
photo-absorptions lead to photo-dissociations \citep{Jura74b}.

We can substitute $R\,n\,n({\rm H})$ in Eqs.~\ref{beta04eq} and
\ref{beta15eq} and finally obtain \citep[see also][]{Cui05}:
\begin{equation}
p_{4,0}\,\beta_0 {{N({\rm H}_2, {\rm J=0)}} \over {N({\rm H}_2)}} + 0.021\,\beta_0=A_{4\rightarrow2} {{N({\rm H}_2, {\rm J=4})} \over {N({\rm H}_2)}}
\label{betaeqf}
\end{equation}
and
\begin{equation}
p_{5,1}\,\beta_1 {{N({\rm H}_2, {\rm J=1)}} \over {N({\rm H}_2)}} + 0.049\,\beta_1=A_{5\rightarrow3} {{N({\rm H}_2, {\rm J=5})} \over {N({\rm H}_2)}}~.
\label{betaeqf2}
\end{equation}
One advantage of
these expressions is that $\beta_{0,1}$ can be determined in each component without 
any assumption to estimate $N$(\HI) in the component.
Values obtained for $\beta_{0,1}$ in the different
components are given in Table~4, where it can be seen that the usual assumption
$\beta_{0}$~=~$\beta_{1}$ is verified.

We find the photo-absorption rate of H$_2$ in the Lyman and Werner bands 
$\beta_0 \in [7.6\times10^{-11},2.5\times10^{-9}]$\,s$^{-1} $ ($\beta_0\simeq
1.1\times10^{-10}$\,s$^{-1}$ for the best-fit value) in component 1; $\beta_0\simeq
1\times10^{-9}$\,s$^{-1}$ in component 2; $\beta_0\simeq
1.2\times10^{-9}$\,s$^{-1}$ in component 12 and $\beta_0\simeq
1.5\times10^{-9}$\,s$^{-1}$ in component 13.  
Much lower values are found for components with higher H$_2$ column
densities: for component 9 ($\beta_0\simeq
1.4\times10^{-11}$\,s$^{-1}$), and for component 10 ($\beta_0\simeq 2.4\times10^{-11}$\,s$^{-1}$). We also derive an
upper-limit on the photo-absorption rate
for components 3 ($\beta_0 \la 2.4\times10^{-9}$\,s$^{-1}$).
These values are spread over about 2 orders of magnitude.

\begin{figure}[!ht]
\begin{center}
\includegraphics[width=\hsize,clip=]{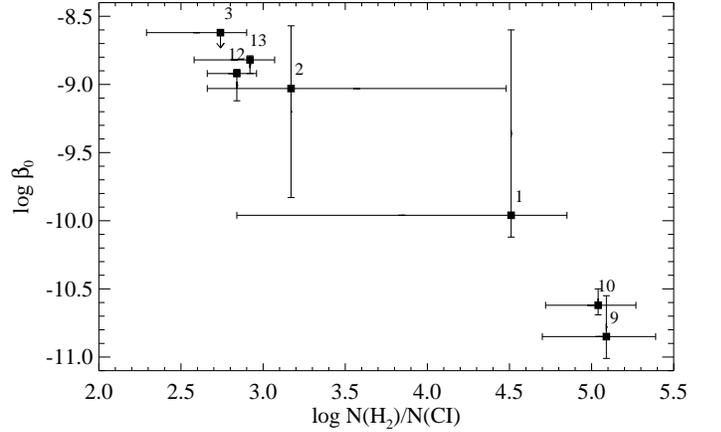}
\caption{\label{beta_H2_CI} Logarithm of the photo-absorption rate as a function of
  $\log N$(H$_2$)/$N$(\CI). Squares stand for the best-fit values. Despite
  large allowed ranges on $\beta_0$ and $N$(H$_2$)/$N$(\CI)
  (dominated by the $N$(H$_2$) uncertainties), there is a clear
  anti-correlation between the two quantities.}
\end{center}
\end{figure}

We plot the photo-absorption rate, $\beta_0$,
versus $\log N$(H$_2$)/$N$(CI) in Fig.~\ref{beta_H2_CI}. 
\citet{Srianand98} suggest that this latter ratio is a good indicator of the physical 
conditions in the absorber as it increases when the ionisation parameter decreases.
There is an anti-correlation (R~$=-0.97$) between $\log \beta_0$ and log~$N$(H$_2$)/$N$(\CI)
that is mostly due to shielding effects.

We can correct for the shielding effect (self-shielding and dust-shielding)
and estimate the UV background intensity outside the cloud.
The photo-dissociation rate is related to the ambient UV flux:

\begin{equation}
R_{\rm diss} = 0.11\,\beta_0 = (4\pi)\,1.1 \times 10^8 J_{\rm LW} S_{\rm shield}\,({\rm s}^{-1})
\end{equation}

\par \noindent where $J_{\rm LW}$ (erg\,s$^{-1}$\,cm$^{-2}$\,Hz$^{-1}$\,sr$^{-1}$) is the UV intensity at
$h\nu = 12.87$~eV, averaged over the solid angle \citep{Abel97, Hirashita05}. 
The term $S_{\rm shield}$ is the 
correction factor for H$_2$ self-shielding \citep{Draine96} and dust extinction. For
$\log N$(H$_2$)~$>14$, it can be expressed as

\begin{equation}
S_{\rm shield}=\left({N({\rm H}_2) \over {\rm 10^{14}\,cm^{-2}}}\right)^{-0.75} e^{-\sigma_{\rm d} N_{\rm d}}~.
\label{shield}
\end{equation}

The term due to dust extinction, $e^{-\sigma_{\rm d} N_{\rm d}}$, is expressed
using the optical depth of the dust in the UV, $\tau_{\rm UV}\equiv\sigma_{\rm d} N_{\rm d}$:

\begin{equation}
\tau_{\rm UV}= 0.879\,{\left({a \over {\rm 0.1 \mu m}}\right)}^{-1} \left({\delta \over {\rm 2\,g\,cm^{-3}}}\right)^{-1} \left({{\cal D}\over{10^{-2}}}\right) \left({N_{\rm H} \over
      {\rm 10^{21}\,cm^{-2}}}\right)
\label{dust_shield}
\end{equation}

\par \noindent where $a$ is the radius of a grain, ${\cal D}$ the dust-to-gas
mass ratio, and $\delta$ the grain material density \citep{Hirashita05}.
Given the low \HI\ column density, the exponential term is always
very close to 1, and $S_{\rm shield}\simeq (N($H$_2) / {10^{14}\,{\rm
    cm}^{-2}})^{-0.75}$; therefore, 
\begin{equation}
 J_{\rm LW} \simeq {{8 \times 10^{-11} \beta_0} \over { S_{\rm shield}}}~.
\label{J_beta_S}
\end{equation}

It can be seen in Table~\ref{beta_table} that for each component, 
the allowed ranges for $J_{\rm LW}$ are all much smaller compared to
the larger allowed
ranges for $\beta_0$. This is especially true for components 1 and 2
(see also Fig.~\ref{beta_H2_CI}) and caused by errors in $N$ being correlated with errors in Doppler parameters. 
Since absorptions from the low J levels are more saturated than those from the higher 
J levels, the derived $N$ are more sensitive to a change in $b$; when $b$ is decreased,
log~$N$(high)/$N$(low) decreases, and the derived value of $\beta_0$ gets lower but the shielding 
correction is larger. 
Therefore the $J_{\rm LW}$ values derived for extreme values of $b$ (see \S\,\ref{C1} and 
Table~\ref{q2348H2t}) are similar.


The derived UV intensity is about one to two orders of
magnitude higher than the typical Galactic UV intensity that is about
$J_{\rm LW,\odot}\simeq 3.2 \times 10^{-20}$\,erg\,s$^{-1}$\,cm$^{-2}$\,Hz$^{-1}$\,sr$^{-1}$ in the
solar vicinity \citep{Habing68}.
Using FUSE, \citet{Tumlinson02} also derive in the Magellanic
  clouds UV fluxes 10 to 100 times higher than in the Milky Way.
This can be interpreted as in-situ star-formation close to
the absorbing clouds. Note that, after shielding correction, the scatter in $J_{\rm LW}$ is much smaller
than the two orders of magnitude variation in $\beta_0$
from one component to the other. 
It must be recalled that this is derived under the
assumption that high-J excitation is mainly due to UV pumping
and formation on dust grains.

\begin{table}
\caption{\label{beta_table} 
UV radiation field and star-formation rate around individual components}
\begin{center}
\begin{tabular}{r c c c c}
\hline
\hline
\# & $\log \beta_0$ & $\log \beta_1$ & $\log J_{\rm LW}$ & $\Sigma_{\rm SFR}$ \\
    &    [s$^{-1}$]  & [s$^{-1}$]    & [erg/s/cm$^{2}$/Hz/sr] & [M$_{\odot}$/yr/kpc$^{2}$]\\
\hline
1  &  $-$9.96$^{+1.36}_{-0.16}$    &  $-$9.96$^{+1.10}_{-0.22}$   & $-$17.02$^{+0.26}_{-0.13}$ & 51$\times 10^{-2}$ \\
2  &  $-$9.00$^{+0.46}_{-0.80}$    &  $-$9.03$^{+0.29}_{-1.21}$   & $-$17.41$^{+0.05}_{-0.14}$ & 21$\times 10^{-2}$  \\
3  &  $\la -$8.62                &  $\la-$8.23               & $\la-$17.74                & $\la$ 9$\times 10^{-2}$  \\
9  &  $-$10.85$^{+0.30}_{-0.16}$   &  $<-$10.67           & $-$18.01$^{+0.35}_{-0.14}$ &  5$\times 10^{-2}$  \\
10 &  $-$10.62$^{+0.12}_{-0.07}$   &  $<-$10.96           &  $-$17.82$^{+0.27}_{-0.02}$ &  8$\times 10^{-2}$  \\
12 &  $-$8.92$^{+0.03}_{-0.20}$    &  $<-9.03$                  & $-$17.70$^{+0.08}_{-0.28}$ & 10$\times 10^{-2}$  \\
13 &  $-$8.82$^{+0.03}_{-0.10}$    &  $<-8.74$                  & $-$17.62$^{+0.37}_{-0.30}$ & 13$\times 10^{-2}$  \\
\hline
\end{tabular}
\end{center}
\end{table}


The observed ratio of the ambient UV flux to that in the solar vicinity, 
$\chi = J_{\rm LW}/J_{\rm LW,\odot}$, is in the range $\chi \sim
30-300$.
This should be considered as upper limits, as no other excitation processes 
than UV pumping and formation on dust is considered here. 
In any case, the observed excitations are still higher than in the Galactic halo or 
the Magellanic clouds. 
We therefore safely estimate that the present UV flux is about one order
of magnitude higher than in the ISM.

Using the relation between the star-formation rate (SFR) per unit surface and
$\chi$,  $\Sigma_{\rm SFR}=1.7\times 10^{-3}\chi$~M$_{\odot}$~yr$^{-1}$~kpc$^{-2}$
\citep{Hirashita05}, we derive a SFR in the different components
in the range, $\Sigma_{\rm SFR}\sim 5 - 50 \times 10^{-2}$~M$_{\odot}$~yr$^{-1}$~kpc$^{-2}$. 
The detailed values are given in Table~\ref{beta_table}.
It can be seen that there is a gradient in the SFR through the DLA
profile. The derived SFR is higher close to component 1. 
%
Note that if the star-formation region is larger than about 1~kpc, the associated
Lyman-$\alpha$ emission should be detectable \citep[e.g][]{Moller04,
  Heinmuller06}.


\subsection{\CI\ fine structure \label{parCI}}

\begin{figure}[!ht]
\begin{center}
\includegraphics[width=\hsize,clip=]{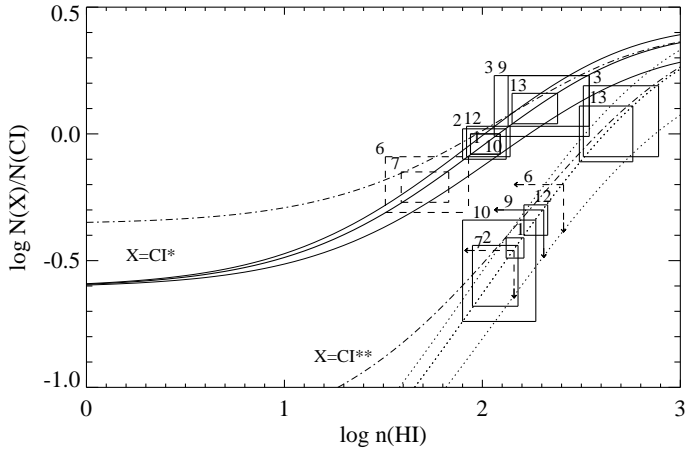}
\caption{\label{CIfs} Analysis of the \CI\ fine-structure relative populations
using results from the POPRATIO code \citep{Silva01}.
Solid and dashed lines give, respectively, $N$(\CI$^*$)/$N$(\CI) and $N$(\CI$^{**}$)/$N$(\CI)
versus gas density for a Galactic ambient flux. The corresponding thick lines stand 
for the model with $T=140$~K, whereas the thin lines stand for $T=200$~K (upper
  curves) and $T=80$~K (lower curves).
The dashed-dotted lines represent the model with
an ambient flux 30 times higher than the Galactic flux.
Boxes show the observed 1\,$\sigma$ range vertically for $N$(\CI$^*$)/$N$(\CI) and 
$N$(\CI$^{**}$)/$N$(\CI) 
and horizontally the corresponding $n$ values for the model with Galactic ambient flux and 
$T$~=~140~K. They are therefore arbitrarily centred on the corresponding curve. 
Note that the results from different models are within errors. The
influence of the UV flux is higher for components 6 and 7, for which
no H$_2$ has been found. The corresponding derived densities are
therefore considered as upper limits.
}
\end{center}
\end{figure}

\begin{figure}
 \begin{center}
\includegraphics[width=\hsize,clip=]{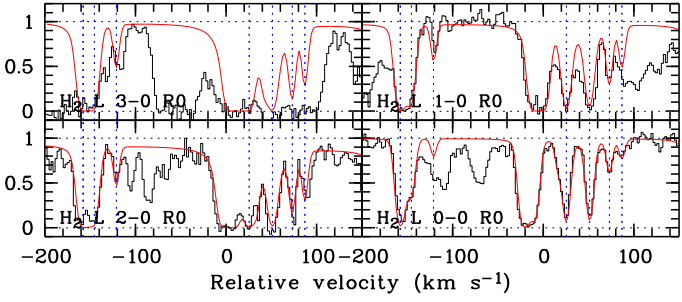}
\caption{\label{J0} H$_2$ rotational level J~=~0. The centre of the
  velocity scale is set at $\zabs=2.4263$. Continuum is seen
  depressed for H$_2$\,L2--0\,R0 and H$_2$\,L1--0\,R0 because of the $\zabs=2.615$
  \lyb\ absorption. Components are marked by dashed lines. The
  strong absorptions at $v\sim$ -10 -- 0 km\,s$^{-1}$  is due to J~=~1
  transition lines.}
 \end{center}
\end{figure}

\begin{figure}
 \begin{center}
\includegraphics[width=\hsize,clip=]{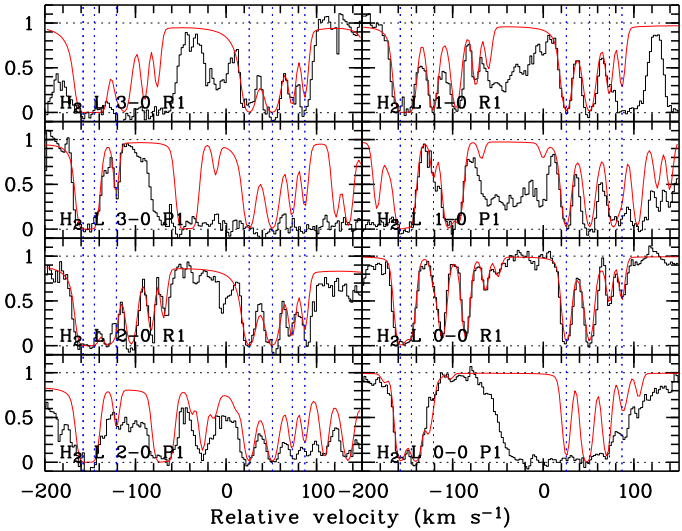}
\caption{\label{J1} H$_2$ rotational level J~=~1. Continuum is
  depressed in panels H$_2$\,L2--0\,R1 to H$_2$\,L1--0\,R1 because of $\zabs=2.615$
  \lyb\ absorption. The J~=~1 components are marked by dashed
  lines, the other ones come from other transitions (in general
  J~=~0 - 2).}
 \end{center}
\end{figure}

\begin{figure}
 \begin{center}
\includegraphics[width=\hsize,clip=]{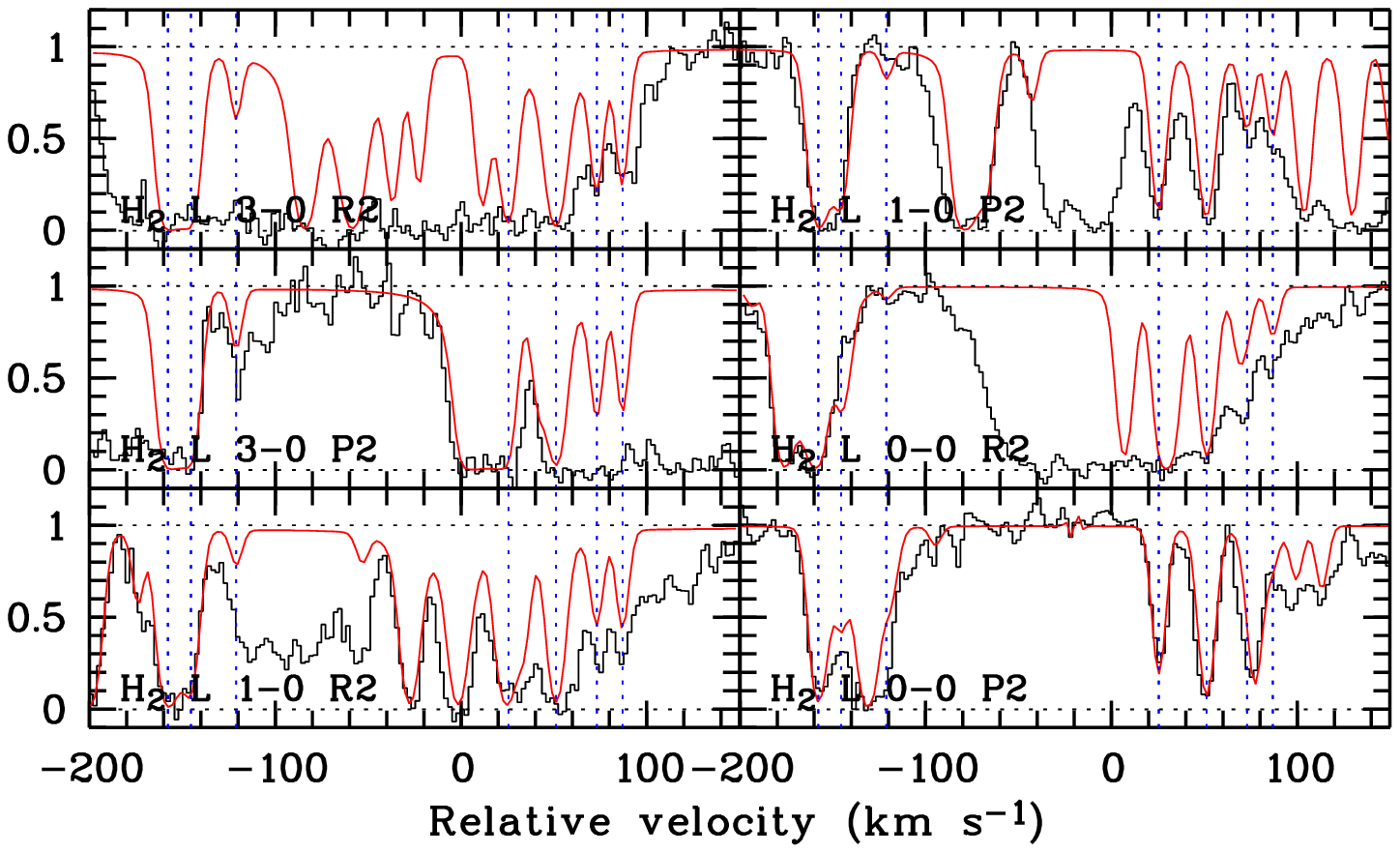}
\caption{\label{J2} H$_2$ rotational level J~=~2. 
H$_2$\,L1--0\,P2 is blended with Fe{\sc ii}$\lambda$1096. }
 \end{center}
\end{figure}

\begin{figure}
 \begin{center}
\includegraphics[width=\hsize,clip=]{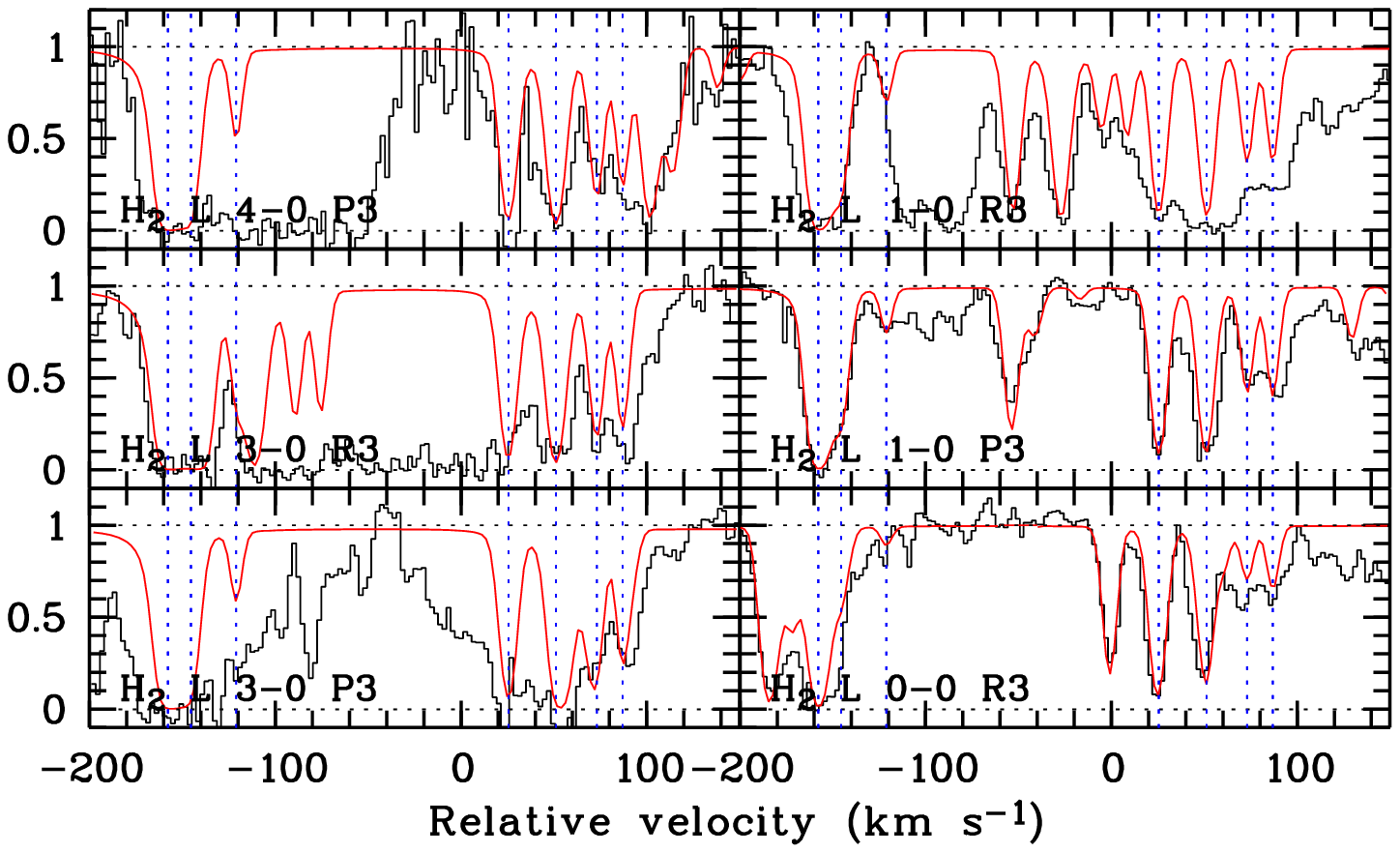}
\caption{\label{J3} H$_2$ rotational level J~=~3. 
H$_2$\,L1--0\,R3 is blended with Fe{\sc ii}$\lambda$1096. }
 \end{center}
\end{figure}

\begin{figure}
 \begin{center}
\includegraphics[width=\hsize,clip=]{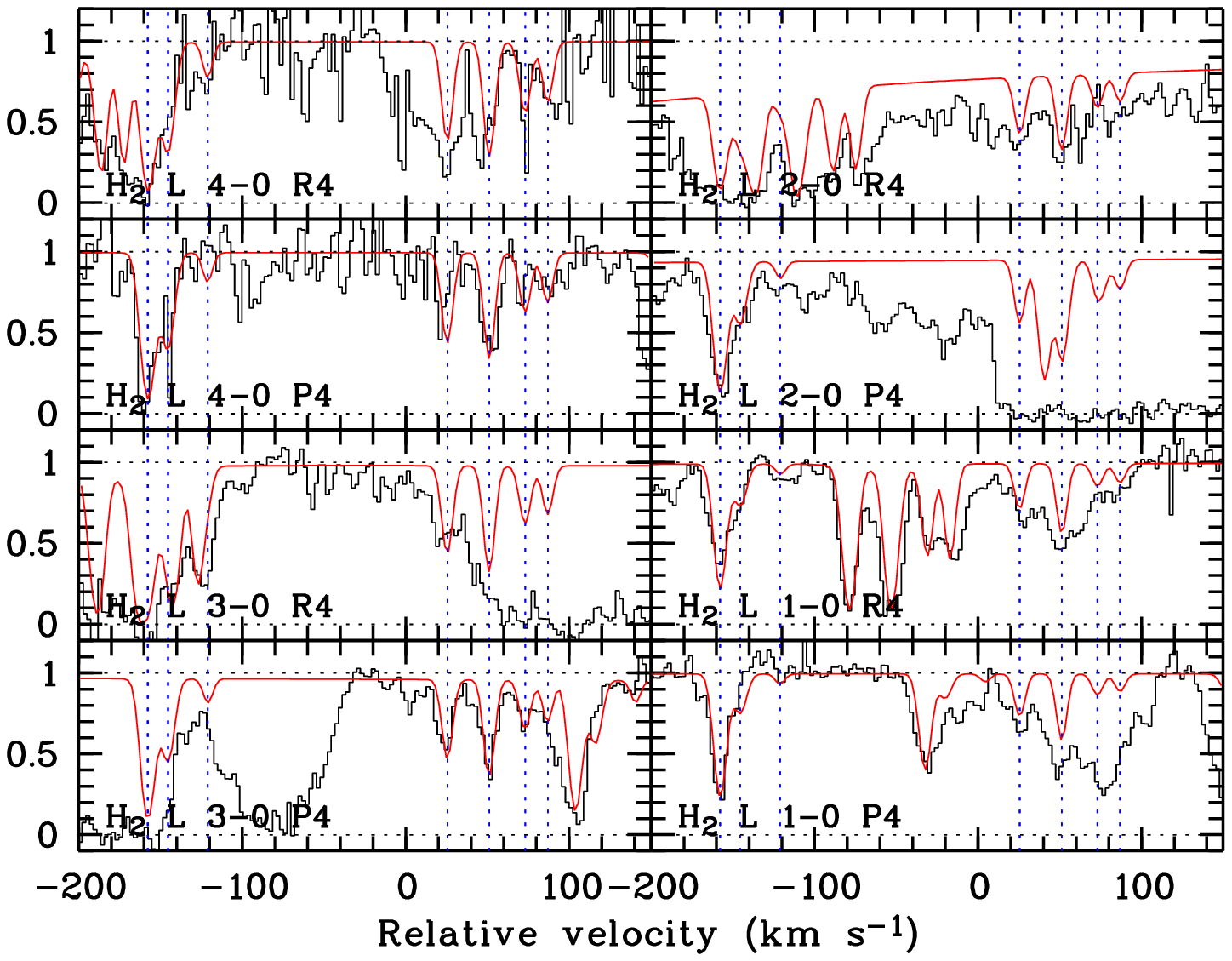}
\caption{\label{J4} H$_2$ rotational level J~=~4. Continuum is
  depressed in some panels by  
  $\zabs\simeq2.6$  \lyb\ absorption. The 
  feature in the H$_2$\,L1--0\,P4 panel at $v \geq -50$ km\,s$^{-1}$ might be due to
  C{\sc iv} absorption at $\zabs\simeq 1.45$.}
 \end{center}
\end{figure}

\begin{figure}
 \begin{center}
\includegraphics[width=\hsize,clip=]{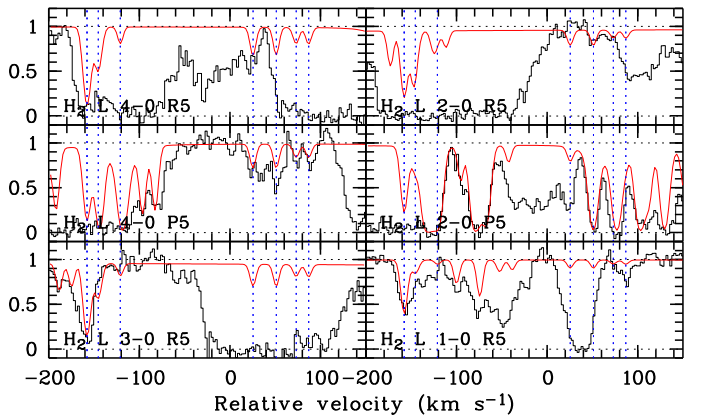}
\caption{\label{J5} H$_2$ rotational level J~=~5. Most lines
are blended with \lya\ or with $z_{\rm
  abs}\simeq 1.45$ C\,{\sc iv} absorptions. 
Column densities are mainly constrained by H$_2$\,L3--0\,R5 for
components 1 \& 2 and by H$_2$\,L4--0\,P5 (resp. H$_2$\,L2--0\,R5) for component 9 (resp. 10).}
 \end{center}
\end{figure}

Neutral carbon is usually seen in components where H$_2$ is detected. 
As the ionisation potential of \CI\ (11.2~eV) 
is similar to the energy of photons in the Werner and Lyman bands, \CI\ is usually 
a good tracer of the physical conditions in the molecular gas.
In the present system, 
\CI\ is detected in all components where H$_2$ is detected but is also detected 
in two additional components (\# 6 and 7). It is possible to derive
constraints on the particle density of the gas from relative populations of the fine-structure 
levels of the \CI\ ground term (2s$^2$2p$^2$\,$^3$P$_{0,1,2}$). In Fig.~\ref{CIfs} 
we compare the ratios $N$(\CI$^*$)/$N$(\CI) and $N$(\CI$^{**}$)/$N$(\CI) with models computed 
with the code POPRATIO \citep{Silva01}. Taking into account the cosmic
microwave background (CMB) radiation with temperature $T_{\rm CMBR}= 2.73\times (1+z)$ and 
assuming an electron density of $n_{\rm e}=10^{-4}\times n_{\rm H}$,
kinetic temperatures in the range $T=80 - 200$~K,
and radiative excitation similar to that in the Milky Way, 
we derive hydrogen
densities of the order of $n$(\HI)~$\simeq 100 - 200$~cm$^{-3}$ for the
components where H$_2$ is detected. It is in turn about $n$(\HI)~$\simeq
50$~cm$^{-3}$
for components without H$_2$ detection if we assume the same kinetic
temperature. If the temperature is higher in these two components (\# 6 and 7), then
densities would be even lower. 
The results are hardly dependent on the exact value of the
ambient 
flux and on the electron and H$_2$ densities for this range of particle 
densities. The most influential processes in that case are the radiative
excitation from the CMB and the collisional excitation depending on
$n$(\HI) and $T$.
To illustrate this, we also assumed an ambient flux 
30 times higher than the 
Galactic flux. The above densities for components 6 and
7 are decreased by a factor of about 2 and even less for the other components. 
Note however, that a very high UV flux ($\sim$ 100 times the Galactic flux) would 
make the observed \CI\
fine-structure column-density ratios inconsistent with the models for the lowest-density components. This again shows that the UV flux derived in the previous
section is probably an upper-limit.

Assuming temperatures between $T\sim 80$~K and $T\sim
200$~K also gives very similar results. Using a Galactic ISM
temperature of $T=80$~K leads to a higher inconsistency between the
$N$(\CI$^{*}$)/$N$(\CI) and $N$(\CI$^{**}$)/$N$(\CI) ratios. This
supports the idea already mentioned in the previous section that 
the kinetic temperature is higher in this system than in the Galactic ISM
(see Figs.~\ref{texfig}, \ref{CIfs} and Table~\ref{q2348n}). Note that, 
because the column densities $N$(\CI$^{**}$) are close to the 
3\,$\sigma$ detection-limit ($\log N($\CI$^{**})=12.50)$ 
and rather dependent on the continuum placement, densities derived from $N$(\CI$^*$)/$N$(\CI) 
are the most reliable.

\begin{table}[!ht]
\caption{\label{q2348n} Densities derived using models in
  Fig.~\ref{CIfs} with $T=140$~K, and the ratios $N$(\CI$^*$)$/N$(\CI)
  (col. 2),  $N$(\CI$^{**}$)$/N$(\CI) (col. 3).}
\begin{center}
\begin{tabular}{r c c c}
\hline
\hline
\#  &  \multicolumn{3}{c}{$n$(\HI) [cm$^{-3}$] } \\
    &  $N$(\CI$^*$)$/N$(\CI) &  $N$(\CI$^{**}$)$/N$(\CI)  &  adopted\,$^1$        \\
\hline
1   &        87--123         &  132--162     & 105$\pm$18              \\
2   &        79--132         &   89--151     & 106$\pm$27              \\
3   &       115--347         &  324--776     & 231$\pm$116              \\
6   &        32--85          &  $<$257       &  $<$59\,$^2$\\
7   &        39--68          &  $<$145       &  $<$54\,$^2$\\
9   &       135--347         &  $<$204       & 170$\pm$35              \\
10  &        87--123         &   79--186     & 105$\pm$18              \\
12  &        83--138         &  162--214     & 111$\pm$28              \\
13  &       141--240         &  309--575     & 191$\pm$50              \\

\hline
\end{tabular}
\end{center}
\footnotesize
$^1$ Because $N$(\CI$^*$) is more reliable than $N$(\CI$^{**}$), we take the corresponding $n$
value as the actual $n$ value (col. 4), unless there is an upper limit
on $N$(\CI$^{**}$) that is then taken into account (component 9).\\
$^2$ The values for components 6 and 7 are considered as upper-limits,
because of the higher dependence on the ambient UV flux.
\normalsize
\end{table}




\section{Discussion and conclusions}
We have detected strong molecular hydrogen absorption lines associated
with the DLA at $z_{\rm abs}$~=~2.4263 toward Q~2348$-$011 in seven 
velocity components spread over $\sim$300~km~s$^{-1}$ and from 
rotational levels J~=~0 up to J~=~5. We used these observations to constrain
the physical conditions in the gas.

From the fine-structure levels of neutral carbon, 
we derived hydrogen volumic densities of the order of 100 to 200~cm$^{-3}$ in the components where
H$_2$ is detected. Densities below 50~cm$^{-3}$ are derived in the two components
where \CI\ is detected but not H$_2$. This agrees with the
results from \citet{Srianand05}. 
The relative populations of the first two rotational levels are all
consistent with a kinetic temperature of about $120$ to $160$~K, under the assumption that $T_{01}$ is a good indicator of
$T_{\rm kin}$. The temperatures derived are higher than typical temperatures in the ISM of the Galaxy.
Detailed models of diffuse molecular clouds should be constructed to 
understand the origin of this difference.
Actual kinetic temperatures could even be slightly higher if the J~=~1 level is
not thermalised \citep[see, e.g.,][]{Lepetit06}, and then densities
would also be slightly lower. 

The physical properties of the H$_2$-bearing gas are characteristic of a cold neutral medium.
Molecular hydrogen in this DLA traces cold gas but with a higher
pressure (of the order of $p\sim 10^{4}$\,cm$^{-3}$K)
than that measured along lines of sight in the Galactic ISM. Temperatures and densities are similar to those observed in the Galactic halo \citep{Wakker06, Richter03}. 

If we assume that \HI\ is distributed over the nine \CI-bearing components proportionally to $N$(\CI), then the mean
\HI\ column density is about 10$^{19.5}$~cm$^{-2}$ in individual
components. The molecular clouds must therefore have an extension along the line 
of sight that is smaller than 1~pc. This supports the view that H$_2$-bearing
gas is clumped into small clouds \citep[see, e.g.,][]{Hirashita03}.

From the relative populations of H$_2$ rotational levels, we
derived photo-absorption rates. Correcting for self-shielding, we are able to estimate the incident UV
flux on the cloud, which we find to be about one 
order of magnitude higher than in the solar vicinity, similar to what is observed through the
  Magellanic clouds \citep{Tumlinson02}. 
This shows that there is ongoing star formation in
this system. 

Searches for the galaxy responsible for the $\zabs=2.4263$ DLA toward
Q\,2348$-$011 have been unsuccessful until now. 
\citet{Mannucci98}, using the NICMOS3 256$^2$ MAGIC camera
at the Calar Alto 3.5~m telescope, reported the detection of an H$\alpha$
emission-line galaxy at $z \simeq 2.43$. The object is located at $\sim 11$ arcsec, 
corresponding to $\sim 90$~kpc for the standard  $\Omega CDM$ cosmology, 
from the line of sight toward Q\,2348$-$011 and the inferred SFR is 78~M$_{\odot}$yr$^{-1}$. 
The emission from the corresponding star-formation region
is not large enough to explain the UV flux in which the gas is embedded.
%
\citet{Bunker99} spectroscopically observed Q\,2348$-$011 over the wavelength range where
possible H$_{\alpha}$ emission at $z$~=~2.43 would be redshifted. From the non-detection
of any emission, they derived an upper limit of about $SFR \leq 21 h^{-2}$~M$_{\odot}$~yr$^{-1}$ 
for the SFR of any possible object exactly aligned with the quasar. 
However, their observations are not very sensitive ($\sim$10$^{-15}$~erg/s/cm$^2$).
%
%
More recently \citet{Colbert02} obtained coronographic H-band images of
the field with the Near-Infrared Camera and Multi-Object Spectrograph mounted on the 
Hubble Space Telescope. With a 5$\sigma$ detection limit $H$~=~22, they detected
three objects at 3.3, 9.7, and 10.7~arcsec or 19, 55, and 61~$h^{-1}$~kpc  from Q\,2348$-$011. 
If at the same redshift as the DLA, these objects are about 1 to 2 magnitudes brighter
than an L$_*$ galaxy. In particular the authors reject the object closer to
the line of sight as a possible DLA galaxy candidate because of its brightness. 

The origin of the UV excess in the vicinity of the gas is therefore unclear.
Either one or several objects within 11~arcsec from the quasar are indeed
at the redshift of the DLA and provide the inferred UV flux through
star-formation activity or the region of star-formation is located on the line
of sight to the quasar that is exactly aligned with it. In that case, Lyman-$\alpha$ emission should be 
observed at the bottom of the DLA trough. The SDSS spectrum of
Q\,2348$-$011 is too noisy
to claim anything in this matter. It will thus be very interesting (i) to determine
the spectroscopic redshifts of the objects that are located close to the line
of sight and (ii) to perform deep intermediate-resolution spectroscopy of the quasar
to search for any Lyman-$\alpha$ emission.



\acknowledgement{We thank the anonymous referee for a thorough reading of
the manuscript and useful comments. PN is supported by a PhD fellowship
from ESO. PPJ and RS gratefully acknowledge support from the 
Indo-French Centre for the Promotion of Advanced Research (Centre Franco-Indien 
pour la Promotion de la Recherche Avanc\'ee) under contract No. 3004-3.
}
\clearpage


\clearpage

\bibliographystyle{aa.bst}
\bibliography{mybib}

\end{document}